\documentclass{emulateapj}
\submitted{Received 2003 December 8; accepted 2004 July 26}
\journalinfo{THE ASTROPHYSICAL JOURNAL, 615, 2004 November 10}
\shortauthors{SAIJO}
\shorttitle{The Collapse of Differentially Rotating SMSs}

\begin{document}
%%%%%%%%%%%%%%%%%%%%%%%%%%%%%%%%%%%%%%%%%%%%%%%%%
% Title page
%%%%%%%%%%%%%%%%%%%%%%%%%%%%%%%%%%%%%%%%%%%%%%%%%
%
%%%%%%%%%%%%%%%%%%%%
% Title
%%%%%%%%%%%%%%%%%%%%
\title
{The Collapse of Differentially Rotating Supermassive Stars:
Conformally Flat Simulations}
%
%%%%%%%%%%%%%%%%%%%%
% Author 1
%%%%%%%%%%%%%%%%%%%%
\author{Motoyuki Saijo}
\email{saijo@tap.scphys.kyoto-u.ac.jp}
%
%%%%%%%%%%%%%%%%%%%%
% Address 1
%%%%%%%%%%%%%%%%%%%%
\affil
{Department of Physics, Kyoto University,
Kyoto 606-8502, Japan}
%
%%%%%%%%%%%%%%%%%%%%%%%%%%%%%%%%%%%%%%%%%%%%%%%%%
\begin{abstract}
%%%%%%%%%%%%%%%%%%%%%%%%%%%%%%%%%%%%%%%%%%%%%%%%%
We investigate the gravitational collapse of rapidly rotating relativistic
supermassive stars by means of a 3+1 hydrodynamical simulations in conformally
flat spacetime of general relativity.  We study the evolution of differentially
rotating supermassive stars of $q \equiv J/M^{2} \sim 1$ ($J$ is the 
angular momentum and $M$ is the gravitational mass of the star) from the 
onset of radial instability at $R/M \sim 65$ ($R$ is the circumferential 
radius of the star) to the point where the conformally flat approximation 
breaks down.  We find 
that the collapse of the star of $q \gtrsim 1$, a radially unstable 
differentially rotating star form a black hole of $q \lesssim 1$.  The main 
reason to prevent formation of a black hole of $q \gtrsim 1$ is that quite 
a large amount of angular momentum stays at the surface.  We also find 
that most of the mass density collapses coherently to form a supermassive 
black hole with no appreciable disk nor bar.  
In the absence of nonaxisymmetric deformation, 
the collapse of differentially rotating supermassive stars from the onset 
of radial instability 
are the promising sources of burst and quasinormal ringing waves in the 
Laser Interferometer Space Antenna.
%%%%%%%%%%%%%%%%%%%%%%%%%%%%%%%%%%%%%%%%%%%%%%%%%
\end{abstract}
%%%%%%%%%%%%%%%%%%%%%%%%%%%%%%%%%%%%%%%%%%%%%%%%%

%%%%%%%%%%%%%%%%%%%%%%%%%%%%%%%%%%%%%%%%%%%%%%%%%
%%%%%%%%%%%%%%%%%%%%%%%%%%%%%%%%%%%%%%%%%%%%%%%%%
%%%
%%%   Section I : Introduction
%%%
%%%%%%%%%%%%%%%%%%%%%%%%%%%%%%%%%%%%%%%%%%%%%%%%%
%%%%%%%%%%%%%%%%%%%%%%%%%%%%%%%%%%%%%%%%%%%%%%%%%
%%%%%%%%%%%%%%%%%%%%
\section{Introduction}
%%%%%%%%%%%%%%%%%%%%
There is increasing evidence that supermassive black holes (SMBHs)
exist at the center of all galaxies, and that they are the sources which
power active galactic nuclei and quasars \citep{Rees98}. 
For example, VLBI observations of the Keplerian disk around an object
in NGC4258 indicate that the central object has a mass $M \sim 2.0
\times 10^{7} M_{\odot}$.  Also,
large numbers of observations are provided by the Hubble space
telescope suggesting that SMBHs exist in galaxies such as M31 ($7.0
\times 10^{7} M_{\odot}$), M87 ($3.4 \times 10^{9} M_{\odot}$)
and our own galaxy ($3.7 \times 10^{6} M_{\odot}$) \citep[see for
example,][for a brief overview]{Kormendy}. 
Although evidence of the existence of SMBHs is compelling, the actual 
formation process of these objects is still uncertain \citep{Rees01}.
Several different scenarios have been proposed, some based on stellar
dynamics, others on gas hydrodynamics, and still others which combine 
the processes. At present, there is no definitive
observation as yet which confirms or rules out any one of these
scenarios.

Here we discuss the collapse of a supermassive star (SMS) as one scenario of 
formation of an SMBH.  This subject is also interesting from a viewpoint of 
general relativity, since the onset of radial instability at the mass shedding 
limit takes place at $J/M^2 \approx 1$ ($J$ is the angular momentum, $M$ is
the total gravitational energy) in the uniformly rotating SMS 
\citep{BS99,Shibata04}, which is also considered as a critical threshold 
between formation of a stationary black hole (BH) and of a naked singularity. 
The BH uniqueness theorem says that 
a complete gravitational collapse of a body always results in a 
BH rather than a naked singularity, and that the final state of the BH should 
go into a stationary Kerr BH \citep[e.g.,][]{Wald}.  However there is a 
candidate to except this cosmic censor.  The collapse of a prolate spheroid 
with large semimajor axis leads to spindle singularities without an apparent 
horizon \citep{NST, ST91, ST92}\footnote{In order to prove violation
of cosmic censor precisely, one might compute a global event horizon instead
of an apparent horizon, which is difficult to treat numerically in dealing 
with a singularity.}.  Therefore, a collapse of a star with 
the critical value $J/M^{2}\sim 1$ may show us an interesting phenomenon in 
general relativity.  What happens when the star
of $J/M^{2} \sim 1$ collapses?  What is the final fate of a collapsing star 
of $J/M^{2} \sim 1$?  Numerical simulations can clearly show the answers to 
these questions.

The gravitational collapse of a star of $J/M^{2} \sim 1$ has been 
investigated most of all in axisymmetric spacetime.
The pioneering study in this field was made by 
\citet{Nakamura81}.  He set up a differentially rotating star, adding radial 
velocity to induce the collapse.  
He found that the criterion of formation of a BH apparent horizon is 
$J/M^{2} \approx 0.95$.  The following study has 
been done by  \citet{SP85, SP86}.  They set up a uniformly rotating $n=1$ 
polytropic star and deplete 99 \% of the whole pressure to induce the 
collapse.  They found that the criterion of BH formation is 
$a_{\rm crit} / M \pm 0.2$, and when the star exceeds the above value, 
flattened disk formed ($a_{\rm crit} / M = 1.2$ for the case of 
99 \% pressure deplete).  Note that $a_{\rm crit} (\equiv J / M)$ 
represents the critical Kerr parameter that the collapse 
proceeds to a BH.
Finally \citet{Shibata00} performed a collapse of a differentially rotating 
$n=1$ polytropic star.  He also depleted the pressure of the star to induce 
the collapse.  He found that when $J/M^{2}$ is less than 0.5, a BH is 
formed when the rest mass is larger than the maximum mass of the pressure 
depleted $J$ constant sequence.  When
$J/M^{2}$ is slightly less than $1$, a BH formed when the 
rest mass is sufficiently larger than the maximum allowed mass on the pressure 
depleted $J$ constant sequence.  He also found, by comparing his results from
polytropic evolution and from $\Gamma$-law evolution of the hydrodynamics, 
that shock heating prevents the prompt collapse to a BH under the 
condition of $J/M^{2} \sim 1$.

At the same time we have submitted our paper, several groups have also 
investigated the collapse of a relativistic star of $J/M^{2} \sim 1$ 
in full general relativity.  \citet{Shibata04} investigated the onset of
radial instability of uniformly rotating stars in 2D and find that the 
criterion of $J/M^{2}$ is  $\gtrsim 1$ in a very soft polytropic
equation of state ($n \approx 3.01$ -- $3.05$; $n$ is a polytropic index).  
\citet{DSY} investigated the collapse of a differentially rotating 
$n=1$ polytropic star in 3D by depleting the pressure and find that 
the criterion of BH formation is $J/M^{2} \approx 1$, and when the 
star exceeds the above value, the collapsing star forms a torus
which fragments into nonaxisymmetric clumps.  \citet{SS04}
studied the collapsing star of $J/M^{2} \sim 1$ in 2D with a polytropic 
index of $n=1$ and $2$ and find that the criterion for BH formation is 
determined by the central $q_{\rm c}$ ($\equiv j/m^{2}$; $j$ is 
the specific angular momentum and $m$ is the cylindrical mass) at the 
initial state of the star.

The purpose of this paper is the following threefold.  The first is to 
verify the nature of the gravitational collapse of a differentially 
rotating radially unstable star from the viewpoint of cosmic censor.  If we 
collapse a star of $J/M^{2} \gtrsim 1$, we expect that the star cannot 
directly form a BH of $J/M^{2} \gtrsim 1$ because of cosmic censor.  
Therefore it is 
important to find the main cause to prevent BH formation of $J/M^{2} 
\gtrsim 1$.  From the previous computational results, shock heating 
\citep{Shibata00} and core bounce \citep{Nakamura81, SP85, SP86, 
Shibata00} are the dominant phenomena to prevent a star from forming a BH.  
However, all of the previous calculations have set up violent initial data 
sets, adding 
radial velocity or depleting pressure, to induce the collapse, it may cause 
abrupt transport of the energy.  We therefore set up a  mild, natural 
situation, that is a collapse of differentially rotating stars from the 
onset 
of radial instability, to focus on a graduate transport of the energy.  Also, 
we compute the collapsing star in 3D to allow shock propagation or bar 
formation, if it occurs. 

The second is to determine the final outcome of the collapse of differentially 
rotating SMSs.  
Two different types of rotation profile arises during the quasi-static 
evolution of the rotating star, that depends on the environment.  One is a
uniformly rotating star, maintained uniform rotation during 
the quasi-static evolution by sufficiently strong viscosity or strong magnetic 
field.  The other is a differentially rotating star, impossible to retain 
uniform rotation in low viscosity and
low magnetic field (even the star initially takes uniform rotation).
For the collapse of a uniformly rotating SMS from the onset of 
radial instability, \citet{SBSS02} studied 3D relativistic hydrodynamic 
simulation in the post-Newtonian gravitation and found that the collapse is 
coherent and that it is likely to form an SMBH with 
no significant bar nor disk formation.   Followup computation has been 
performed in 2D hydrodynamics in full general relativity and found that
approximately 10 \% of the total mass can form a disk, while approximately 
90 \% of that should form a BH \citep{SS02}.  
For the differentially rotating stars, the final outcome may strongly depends 
on the amount and the distribution of initial angular momentum, and 
the nature of angular momentum distribution of SMSs 
is largely unknown.  
We also do not know what path does a differentially rotating SMS take
during the quasi-static evolution.  When the amount of angular momentum 
of the star is sufficient, the star seems to follow in a quasi-static 
evolution up 
to the mass-shedding limit.  \citet{NS} find in the Newtonian quasi-static 
evolution that bar formation is inevitable before reaching mass-shedding 
limit.  When the amount of angular momentum of the star is not sufficient,
the star seems to collapse at the onset of radial instability before
reaching secular instability ($T/W \approx 0.14$ for a uniformly rotating
incompressible Newtonian stars \citep{Chandra69}) or mass-shedding limit.
The purpose of this paper is to examine the collapse of the SMS 
from the onset of radial instability.  When we consider the 
collapse of radially unstable differentially rotating 
stars, the final outcome has a possibility of differing from the 
collapse of uniform rotation because of the strong centrifugal force at the 
central core, which may prevent the prompt collapse.  What is the final fate 
of the collapse?  Does the star fragment due to the growth of the degree of 
differential rotation?  Does the disk form during the collapse?  Relativistic 
simulation can clearly answer to these questions.

Finally, it is important to probe whether the collapse of differentially 
rotating SMSs could be a promising source of gravitational waves.
Direct detection of gravitational waves by ground based and space based
interferometers is of great importance in general relativity, in astrophysics, 
and in cosmology \citep[e.g.,][]{Thorne98}.  The catastrophic collapse is one 
of the promising sources of gravitational waves \citep[e.g.,][]{New03}.  
For a gravitational collapse of the star in a dynamical timescale, there 
are two main reasons that prevent a prompt collapse of the star, which 
should produce gravitational waves at that moment.  One reason is core bounce 
and/or shock heating.  Suppose gravitational force is balanced to the 
centrifugal force ($M/R^{2} \sim R \Omega^{2}$) in Newtonian gravity. 
The total mass and the angular momentum ($J \sim M R^{2} \Omega$)
are almost conserved during the collapse, assuming that
gravitational radiation takes a little role during the collapse in the 
absence of nonaxisymmetric deformation prior to form a BH.  Note that  
$\Omega$ is angular velocity, $R$ is the radius.  We may estimate the 
bounce radius of the star in a global sense as
$R_{\rm bounce} \approx M ( J/M^{2} )^{2}$.  Since the horizon radius is 
roughly the order of $M$, core bounce might take place for the case 
$J/M^{2} \gtrsim 1$, which prevents violation of cosmic
censor \citep{Nakamura81}.  The other is bar formation.
From the dimensional analysis, we can describe the rotational kinetic energy
$T$ and the gravitational binding energy $W$ as 
$T \sim M R^{2} \Omega^{2}$ and $W \sim M^{2} / R$.  We also accept the 
assumption that total mass and angular momentum are almost conserved 
during the collapse.  We could estimate the radius of bar formation in 
terms of the ratio of the rotational kinetic energy to the gravitational 
binding energy 
as $R_{\rm bar} \approx (M/R) ( T/W )^{-1} ( J/M^{2} )^{2}$.
Since the dynamical instability for a uniformly rotating, incompressible
Newtonian star sets in at $T/W \sim 0.27$ \citep{Chandra69} and for 
relativistic gravitation as $T/W \sim 0.24 - 0.26$ \citep{SBS00, SSBS01}, 
bar formation takes place at the radius $R \sim 4 M$ for the collapse of 
$J/M^{2} \sim 1$.  Therefore, we may expect that a gravitational 
collapse of $J/M^{2} \sim 1$ is a promising source of quasi-periodic 
gravitational waves.

This paper is organized as follows.  In \S~\ref{sec:CF} we present 
basic equations of our conformally flat approximation in general relativity.  
We demonstrate our code tests in \S~\ref{sec:ctest}.  We discuss our 
initial data sets and our numerical results for our catastrophic collapse in 
\S~\ref{sec:jm2}.   In \S~\ref{sec:Discussion} we summarize our findings.  
Throughout this paper, we use geometrized units ($G=c=1$) and adopt Cartesian
coordinates $(x,y,z)$ with the coordinate time $t$.  Greek and Latin
indices run over $(t, x, y, z)$ and $(x, y, z)$, respectively.

%%%%%%%%%%%%%%%%%%%%%%%%%%%%%%%%%%%%%%%%%%%%%%%%%
%%%%%%%%%%%%%%%%%%%%%%%%%%%%%%%%%%%%%%%%%%%%%%%%%
%%%
%%%   Section II : 3+1 Relativistic Hydrodynamics in Conformally Flat Spacetime
%%%
%%%%%%%%%%%%%%%%%%%%%%%%%%%%%%%%%%%%%%%%%%%%%%%%%
%%%%%%%%%%%%%%%%%%%%%%%%%%%%%%%%%%%%%%%%%%%%%%%%%
%%%%%%%%%%%%%%%%%%%%
\section{3+1 Relativistic Hydrodynamics in Conformally Flat Spacetime}
\label{sec:CF}
%%%%%%%%%%%%%%%%%%%%

In this section, we briefly describe the conformally flat spacetime 
\citep[e.g.,][]{IN}.
We solve the fully relativistic equations of hydrodynamics, but
neglect nondiagonal spatial metric components \citep{WM89,WM95}.

%%%%%%%%%%%%%%%%%%%%%%%%%%%%%%%%%%%%%%%%%%%%%%%%%
\subsection{The gravitational field equations}

We define the spatial projection tensor $\gamma^{\mu\nu} \equiv
g^{\mu\nu} + n^{\mu} n^{\nu}$, where $g^{\mu\nu}$ is the spacetime
metric, $n^{\mu} = (1/\alpha, -\beta^i/\alpha)$ the unit normal to a
spatial hypersurface, and where $\alpha$ and $\beta^i$ are the lapse
and shift.  Within a first post-Newtonian approximation, the spatial
metric $g_{ij} = \gamma_{ij}$ may always be chosen to be conformally flat
%%%%%%%%%%
\begin{eqnarray}
\gamma_{ij} = \psi^{4} \delta_{ij},
\end{eqnarray}
%%%%%%%%%%
where $\psi$ is the conformal factor \citep[see][]{Chandra65, BDS}.
The spacetime line element then reduces to
%%%%%%%%%%
\begin{eqnarray}
ds^{2} &=&
( - \alpha^{2} + \beta_{k} \beta^{k} ) dt^{2} + 2 \beta_{i} dx^{i} dt 
+ \psi^{4} \delta_{ij} dx^{i} dx^{j}.
\end{eqnarray}
%%%%%%%%%%
We adopt maximal slicing, for which the trace of the extrinsic
curvature $K_{ij}$ vanishes,
%%%%%%%%%%
\begin{equation}
K \equiv \gamma^{ij} K_{ij} = 0.
\end{equation}  
%%%%%%%%%%
The gravitational field equations in conformally flat spacetime for the 
five unknowns $\alpha$, $\beta^i$, and $\psi$ can then be derived conveniently 
from the 3+1  formalism \citep[e.g.,][]{SSBS01}.

Since the spatial metric is conformally flat, the transverse part of
its time derivative vanishes.  The transverse part of the evolution
equation of the spatial metric therefore relates the extrinsic
curvature to the shift vector,
%%%%%%%%%%
\begin{equation}
2 \alpha \psi^{-4} K_{ij} =
\delta_{jl} \partial_{i} \beta^{l} + \delta_{il} \partial_{j} \beta^{l} - 
\frac{2}{3} \delta_{ij} \partial_{l} \beta^{l}.
\label{eqn:EvolutionMetric}
\end{equation}
%%%%%%%%%%
Inserting eq. (\ref{eqn:EvolutionMetric}) into the momentum constraint
equation, then, yields an equation for the shift $\beta^{i}$
%%%%%%%%%%
\begin{eqnarray}
\label{eqn:MomentumC}
&&
\delta_{il} \triangle \beta^{l} + 
\frac{1}{3} \partial_{i} \partial_{l} \beta^{l} =16 \pi \alpha J_{i} +
\left(\partial_{j}\ln \left( \frac{\alpha}{\psi^{6}} \right)\right)
\nonumber \\
&& \hspace{1cm} 
\times
\left( 
   \partial_{i} \beta^{j} + \delta_{il} \delta^{jk} \partial_{k} \beta^{l} -
  \frac{2}{3} \delta_{i}^{j} \partial_{l} \beta^{l} 
\right),
\end{eqnarray}
%%%%%%%%%%
where $\Delta \equiv \delta^{ij}\partial_i\partial_j$ is the flat
space Laplacian and $J_{i} \equiv -n^{\mu} \gamma^{\nu}_{~i} T_{\mu
\nu}$ is the momentum density.  In the definition of $J_{i}$,
$T_{\mu\nu}$ is the stress energy tensor.

The conformal factor $\psi$ is determined from the Hamiltonian constraint
%%%%%%%%%%
\begin{equation}
\triangle \psi =
- 2 \pi \psi^{5} \rho_{\rm H} - \frac{1}{8} \psi^{5} K_{ij} K^{ij},
\label{eqn:HamiltonianC}
\end{equation}
%%%%%%%%%%
where $\rho_{\rm H} \equiv n^{\mu} n^{\nu} T_{\mu \nu}$ is the mass-energy
density measured by a normal observer.

Maximal slicing implies $\partial_t K = 0$, so that the trace of the 
evolution equation for the intrinsic curvature yields an equation for 
the lapse $\alpha$
%%%%%%%%%%
\begin{equation}
\triangle (\alpha \psi) = 2 \pi \alpha \psi^{5} 
(\rho_{\rm H} + 2 S) + \frac{7}{8} \alpha \psi^{5} K_{ij} K^{ij},
\label{eqn:evolution}
\end{equation}
%%%%%%%%%%
combined with eq. (\ref{eqn:HamiltonianC}), where $S  = \gamma_{jk} T^{jk}$.

Therefore, conformally flat gravitational field equations for the five 
unknowns $\psi$, $\alpha \psi$, $\beta^{i}$ can be derived by eqs.
(\ref{eqn:MomentumC}) - (\ref{eqn:evolution}).

%%%%%%%%%%%%%%%%%%%%%%%%%%%%%%%%%%%%%%%%%%%%%%%%%
\subsection{The matter equations}
For a perfect fluid, the energy momentum tensor takes the form 
%%%%%%%%%%
\begin{equation}
T^{\mu \nu} = 
\rho \left( 1 + \varepsilon + \frac{P}{\rho} \right) u^{\mu} u^{\nu} +
Pg^{\mu\nu},
\end{equation}
%%%%%%%%%%
where $\rho$ is the rest-mass density, $\varepsilon$ the specific
internal energy, $P$ the pressure, and $u^{\mu}$ the four-velocity.

We adopt a $\Gamma$-law equation of state in the form
%%%%%%%%%%
\begin{equation}
P = (\Gamma - 1) \rho \varepsilon,
\label{gammalaw1}
\end{equation}
%%%%%%%%%%
where $\Gamma$ is the adiabatic index which we set 
$\approx 1.33 \sim 1.34$ in this paper.

In the absence of thermal dissipation, eq.~(\ref{gammalaw1}), together
with the first law of thermodynamics, implies a polytropic equation of
state
%%%%%%%%%%
\begin{equation}
P = \kappa \rho^{1+1/n}, \label{gammalaw}
\end{equation}
%%%%%%%%%%
where $n=1/(\Gamma-1)$ is the polytropic index and $\kappa$ is a
constant.  

From $\nabla_{\mu} T^{\mu\nu}=0$ together with the equation of state
(eq. [\ref{gammalaw1}]), we can derive the energy and Euler equations
according to
%%%%%%%%%%
\begin{eqnarray}
&&
\frac{\partial e_{*}}{\partial t}+
\frac{\partial (e_{*} v^{j})}{\partial x^{j}} = 
- \frac{1}{\Gamma}(\rho \epsilon)^{-1+1/\Gamma} 
P_{\rm vis}
\frac{\partial}{\partial x^{i}} 
( \alpha u^{t} \psi^{6} v^{i} )
\label{eqn:Energy}
,\\
&&
\frac{\partial(\rho_{*} \tilde u_{i})}{\partial t}
+ \frac{\partial (\rho_* \tilde u_{i} v^{j})}{\partial x^{j}} 
=
- \alpha \psi^{6} (P + P_{\rm vis})_{,i} 
- \rho_{*} \alpha \tilde u^{t} \alpha_{,i} 
\nonumber \\
&& \hspace{2cm}
+ \rho_{*} \tilde u_{j} \beta^{j}_{~,i}
+ \frac{2 \rho_{*} \tilde u_{k} \tilde u_{k}}{\psi^{5} \tilde u^{t}} 
\psi_{,i}
,  
\label{eqn:Euler}
\end{eqnarray}
%%%%%%%%%%
where 
%%%%%%%%%%
\begin{eqnarray}
e_{*} &=& (\rho \varepsilon)^{1/\Gamma} \alpha u^{t} \psi^{6},\\
v^{i} &=& {dx^i \over dt}=\frac{u^{i}}{u^{t}},\\
\rho_{*} &=& \rho \alpha u^{t} \psi^{6},\\
\tilde{u}^{t} &=& ( 1 + \Gamma \varepsilon ) u^{t}
,\\
\tilde{u}_{i} &=& ( 1 + \Gamma \varepsilon ) u_{i}
,
\end{eqnarray}
%%%%%%%%%%
and $v^{i}$, $P_{\rm vis}$ is the 3-velocity, 
pressure viscosity, respectively.  Note that we treat the matter fully 
relativistically; the conformally flat approximation only enters through 
simplifications in the coupling to the gravitational fields.  Note also that 
we include an artificial viscosity, since core bounce might occur in our 
gravitational collapse and that should produce shocks.  We will explain our 
form of an artificial viscosity in \S~\ref{sec:ctest}. As a 
consequence to treat shocks we also need to solve the continuity equation
%%%%%%%%%%
\begin{equation}
\frac{\partial \rho_{*}}{\partial t}
+\frac{\partial (\rho_{*} v^{i})}{\partial x^{i}} = 0,
\label{eqn:BConservation}
\end{equation}
%%%%%%%%%%
separately.

We solve the matter evolution in second order accurate in space and time with the transfer
scheme of \citet{vanLeer, ON89}.

%%%%%%%%%%%%%%%%%%%%%%%%%%%%%%%%%%%%%%%%%%%%%%%%%
%%%%% Techniques %%%%%
\subsection{Numerical techniques for solving gravitational field equations}
\label{subsec:NT}

There are three key issues to solve gravitational field equations numerically.
The first is to introduce symmetric tensor $\hat{A}_{ij}$ related to the extrinsic
curvature $K_{ij}$ as \citep{SN95}
%%%%%%%%%%
\begin{eqnarray}
\hat{A}_{ij} \equiv \psi^{2} \left( K_{ij} - \frac{1}{3} \gamma_{ij} K \right),
\\
\hat{A}^{ij} \equiv \psi^{10} \left( K^{ij} - \frac{1}{3} \gamma^{ij} K \right).
\end{eqnarray}  
%%%%%%%%%%
Therefore, we describe the equations of 
conformal factor (eq. [\ref{eqn:HamiltonianC}]) and 
lapse (eq. [\ref{eqn:evolution}]) as
%%%%%%%%%%
\begin{eqnarray}
\triangle \psi &=& 
- 2 \pi \psi^{5} \rho_{\rm H} - 
\frac{1}{8} \psi^{-7} \hat{A}_{ij} \hat{A}^{ij}
\label{eqn:CFactor}
,\\
\triangle (\alpha \psi) &=&
2 \pi \alpha \psi (\rho_{\rm H} + 2 S) +
\frac{7}{8} \alpha \psi^{-7} \hat{A}_{ij} \hat{A}^{ij}
.
\end{eqnarray}
%%%%%%%%%%
Note that the dependence value (such as $\psi$, $\alpha \psi$) of the 
source term is always lower than O($\psi^{1}$), O($(\alpha \psi)^{1}$), 
which is safe for a convergence 
during the iteration.\footnote{Since we compute the variabilities of the matter as $\rho_{*},
e_{*}, \tilde{u}_{i}$, the first term of eq. (\ref{eqn:CFactor}) behaves as 
$-2\pi \psi^{-1} (\psi^{6} \rho_{\rm H})$.}

The second is to decompose the vector type elliptic equation.
The equation for the shift,
%%%%%%%%%%
\begin{eqnarray}
\delta_{ij} \triangle \beta_{j} + 
\frac{1}{3} \partial_{i} \partial_{l} \beta^{l} &=&
\frac{2}{\psi^{6}} \left( \partial^{j} \alpha - 
6 \frac{\alpha}{\psi} \partial^{j} \psi \right)
\hat{A}_{ij} + 16 \pi \alpha J_{i}
\nonumber \\
&\equiv& \hat{J}_{i},
\label{eqn:shifteq}
\end{eqnarray}
%%%%%%%%%%
can be further simplified by introducing a vector $P_{i}$ and a scalar 
$\eta$ according to \citep{Shibata97}
%%%%%%%%%%
\begin{eqnarray}
\triangle P_{i} &=& \hat{J}_{i}
,\\
\triangle \eta &=& -\hat{J}_{i} x^{i}
.
\end{eqnarray}
%%%%%%%%%%
The shift can then be computed from
%%%%%%%%%%
\begin{equation}
\delta_{ij} \beta^{j} = 
\frac{7}{8} P_{i} - \frac{1}{8} 
(\partial_{i} \eta + x^{k} \partial_{i} P_{k})
,
\end{equation}
%%%%%%%%%%
and will automatically satisfy eq.~(\ref{eqn:shifteq}).  
Next, we decompose the symmetric tensor $\hat{A}_{ij}$ as \citep[e.g.,][]{York}
%%%%%%%%%%
\begin{equation}
\hat{A}_{ij} = \hat{A}^{*}_{ij} + (\hat{l} W)_{ij},
\end{equation}
%%%%%%%%%%
where 
%%%%%%%%%%
\begin{eqnarray}
&&
\partial_{j} \hat{A}^{*ij} = \tilde{\gamma}_{ij}
\hat{A}^{*ij} = 0
,\\
&&
(\hat{l} W)_{ij} = \partial_{i} W_{j} + \partial_{j} W_{i} -
\frac{2}{3} \delta_{ij} \partial_{k} W^{k} 
.
\end{eqnarray}
%%%%%%%%%%
Therefore, we can rewrite the Momentum constraint as
%%%%%%%%%%
\begin{equation}
\triangle W_{i} + \frac{1}{3} \partial_{i} \partial_{j} W^{j}
=
8 \pi \psi^{6} J_{i}
.
\end{equation}
%%%%%%%%%%
We also decompose the vector $W^{i}$ with the same manner as the shift;
%%%%%%%%%%
\begin{equation}
\delta_{ij} W^{j} = 
\frac{7}{8} B_{i} - \frac{1}{8} 
(\partial_{i} \chi + x^{k} \partial_{i} B_{k})
,
\end{equation}
%%%%%%%%%%
where $B_{i}$ and $\chi$ satisfies
%%%%%%%%%%
\begin{eqnarray}
\triangle B_{i} &=& 8 \pi \psi^{6} J_{i}
\label{eqn:WB}
,\\
\triangle \chi &=& -8 \pi \psi^{6} J_{i} x^{i}
\label{eqn:Wchi}
.
\end{eqnarray}
%%%%%%%%%%
Note that $J_{i}$ only appears in the presence of matter, which means that
the source terms of eqs. (\ref{eqn:WB}) and (\ref{eqn:Wchi}) are compact.
Therefore, we can compute these values quite accurately. 

To summarize, we have reduced Einstein equations in a conformally flat 
spacetime to 10 elliptic equations for 10 variables 
($B_{i}$, $\chi$, $\psi$, $\alpha \psi$, $P_{i}$, $\eta$),
%%%%%%%%%%
\begin{eqnarray}
\Delta B_{i} &=& 8 \pi \psi^{6} J_{i} \equiv 4 \pi S_{B_{i}},
\label{eqn:CFBx}
\\
\Delta \chi  &=& - 8 \pi \psi^{6} J_{i} x^{i} \equiv 4 \pi S_{\chi},
\\
\Delta \psi  &=& 
- 2 \pi \psi^{5} \rho_{\rm H} - 
\frac{1}{8} \psi^{-7} \hat{A}_{ij} \hat{A}^{ij}
\equiv 4 \pi S_{\psi},
\\
\Delta (\alpha \psi) &=& 
2 \pi \alpha \psi (\rho_{\rm H} + 2 S) +
\frac{7}{8} \alpha \psi^{-7} \hat{A}_{ij} \hat{A}^{ij}
\nonumber \\
& \equiv &
4\pi S_{\alpha\psi}, 
\\
\Delta P_{i} &=& 4 \pi \alpha \hat{J}_{i} \equiv 4 \pi S_{P_{i}},
\\
\Delta \eta  &=& -4 \pi \alpha \hat{J}_{i} x^{i} \equiv 4 \pi S_{\eta}
\label{eqn:CFeta}
.
\end{eqnarray}
%%%%%%%%%%
These Poisson-type equations are solved by imposing the following 
boundary condition at outer boundaries
%%%%%%%%%%
\begin{eqnarray}
B_{x} &=& 
- \frac{x}{r^{3}} \int S_{B_{x}} x d^3 x - 
\frac{y}{r^{3}} \int S_{B_{x}} y d^3 x 
%\nonumber \\
%& & 
+ O(r^{-4}), \\
B_{y} &=& 
- \frac{x}{r^{3}} \int S_{B_{y}} x d^3 x - 
\frac{y}{r^{3}} \int S_{B_{y}} y d^3 x  
%\nonumber \\
%&& 
+ O(r^{-4}) ,\\
B_{z} &=& 
- \frac{z}{r^{3}} \int S_{B_{z}} z d^3 x + O(r^{-4}) ,\\
\chi &=& \frac{1}{r} \int S_{\chi} x^{i} d^3 x + O(r^{-3}),\\
\psi &= &
1 - \frac{1}{r} \int S_{\psi} d^3x + O(r^{-3}),\\
\alpha\psi &=&
1 - \frac{1}{r} \int S_{\alpha\psi} d^3 x + O(r^{-3}),\\
P_{x} &=& 
- \frac{x}{r^{3}} \int S_{P_{x}} x d^3 x - 
\frac{y}{r^{3}} \int S_{P_{x}} y d^3 x 
+ O(r^{-4}), \\
P_{y} &=& 
- \frac{x}{r^{3}} \int S_{P_{y}} x d^3 x - 
\frac{y}{r^{3}} \int S_{P_{y}} y d^3 x  
+ O(r^{-4}) ,\\
P_{z} &=& 
- \frac{z}{r^{3}} \int S_{P_{z}} z d^3 x + O(r^{-4}) ,\\
\eta &=& 
\frac{1}{r} \int S_{\eta} x^{i} d^3 x + O(r^{-3}).
\end{eqnarray}
%%%%%%%%%%

Here we briefly explain our order of computing 10 Poisson-type equations.
First, we solve $B_{i}$ and $\chi$ since the source terms do not depend
on other unknown spacetime variables.  Once we derive the symmetric tensor
$A_{ij}$ from $B_{i}$ and $\chi$, we solve $\psi$ iteratively till the 
convergence.  After that 
we solve $\alpha \psi$ iteratively till the convergence. Finally, we
solve $P_{i}$ and $\chi$.  We use Modified Incomplete Cholesky
decomposition Conjugate Gradient (MICCG) method \citep{MNK} to solve elliptic
equations.

Finally, we should maintain our grid resolution at the central core of the
collapsing star.  The nested grid is one of the most 
appropriate way to handle this problem \citep[see, e.g.,][]{Ruffert92}.  
Our method is a mimic version 
of the nested grid to handle the collapsing star problem, which is based on 
the rigridding method of \citet{SS02}.  For the post homologous collapse 
of the SMS, there are two different timescales in the collapsing star;
one is the collapsing timescale at the center and the other is the one 
at the envelope, which is sufficiently longer than that at the center.  
Therefore, \citet{SS02} add the grid number at the intermediate stage of 
the collapse in order to control both regions, center and envelope of the 
star.  Since we perform the collapse of an SMS in 3D, we concentrate on 
the central core of the collapse due to the limitation of computational 
resource.  In order to maintain spatial grid resolution 
especially at the center, we shrink the total grid to one half as the core 
radius of the star becomes half from the beginning of the present grid 
resolution.  Although this method changes the boundary condition when we 
``zoom in'' the computational domain, it is approximately good when the 
difference of the gravitational mass between the end of the previous 
computational grid and the beginning is small.  In fact, the difference of 
the gravitational mass between every two computational domains is less 
than $\approx 10^{-4}$, and therefore this ``zoom-in'' method should behave
as a 
reasonable approximation for a collapse of differentially rotating SMSs.

We monitor the gravitational mass $M$ and the angular momentum $J$ 
%%%%%%%%%%
\begin{eqnarray}
M &=& - \frac{1}{2 \pi} \oint_{\infty} \nabla^{i} \psi dS_{i}
\nonumber \\
&=&
\int 
\left[ 
  \left[ 
    ( \rho + \rho \varepsilon + P ) (\alpha u^{t})^{2} - P 
  \right] \psi^{5} 
\right.
\nonumber \\
&& \left.
+ \frac{1}{16 \pi} \psi^{5} K_{ij} K^{ij}
\right] d^{3} x 
,\\
J &=&
- \frac{1}{2 \pi} \oint_{\infty} ( x K^{i}_{y} - y K^{i}_{x} ) \psi^{6} dS_{i}
\nonumber \\
&=&
\int (x J_{y} - y J_{x}) \psi^{6} d^{3} x
,
\end{eqnarray}
%%%%%%%%%%
during the evolution.  
In all cases reported in \S~\ref{sec:jm2}, the gravitational mass was
conserved up to $\sim 0.1$\%, and the angular momentum up to $\sim 1$\%
of their initial values.

We also compute proper mass $M_{p}$, rotational kinetic energy $T$, 
gravitational binding energy $W$ in equilibrium as
%%%%%%%%%%
\begin{eqnarray}
M_{p} &=&
\int \rho u^{t} ( 1 + \epsilon ) \sqrt{-g} d^{3} x
%\nonumber \\
%&=&
= \int \rho_{*} ( 1 + \varepsilon ) d^{3} x
,\\
T &=&
\frac{1}{2} \int \Omega T^{t}_{\phi} \sqrt{-g} d^{3} x 
\nonumber \\
&=& \frac{1}{2} \int \Omega (x J_{y} - y J_{x}) \psi^{6} d^{3} x
,\\
W &=&
M_{p} + T - M,
\end{eqnarray}
%%%%%%%%%%
where $\Omega$ the angular velocity of the star.

Since we use a polytropic equation of state at $t=0$, it is convenient to
rescale all quantities with respect to $\kappa$ \citep{CST92}.  Since 
$\kappa^{n/2}$
has dimensions of length, we introduce the following nondimensional
variables
%%%%%%%%%%
\begin{equation}
\begin{array}{c c c}
\bar{t} = \kappa^{-n/2} t
, &
\bar{x} = \kappa^{-n/2} x
, &
\bar{y} = \kappa^{-n/2} y
, \\
\bar{z} = \kappa^{-n/2} z
, &
\bar{\Omega} = \kappa^{n/2} \Omega
, &
\bar{M} = \kappa^{-n/2} M
, \\
\bar{R} = \kappa^{-n/2} R
, &
\bar{J} = \kappa^{-n} J
. &
\end{array}
\end{equation}
%%%%%%%%%%
Henceforth, we adopt nondimensional quantities, but omit the bars for
convenience (equivalently, we set $\kappa = 1$).

%%%%%%%%%%%%%%%%%%%%%%%%%%%%%%%%%%%%%%%%%%%%%%%%%
%%%%%%%%%%%%%%%%%%%%%%%%%%%%%%%%%%%%%%%%%%%%%%%%%
%%%
%%%   Section III : Code tests
%%%
%%%%%%%%%%%%%%%%%%%%%%%%%%%%%%%%%%%%%%%%%%%%%%%%%
%%%%%%%%%%%%%%%%%%%%%%%%%%%%%%%%%%%%%%%%%%%%%%%%%
%%%%%%%%%%%%%%%%%%%%
\section{Code tests}
\label{sec:ctest}
%%%%%%%%%%%%%%%%%%%%

%%%%% Wall Shock Problem %%%%%
First, we demonstrate 1D relativistic wall shock problem to 
check whether our code has an ability to treat shock.  We set up the form of 
artificial pressure viscosity \citep{HSW} as
%%%%%%%%%%
\begin{equation}
P_{\rm vis} =
\cases{ 
C_{\rm vis} 
\rho^{*} ( 1 + \Gamma \epsilon ) (\delta v)^{2},
& for $\delta v \leq 0$;
\cr
0, & for $\delta v \geq 0$,\cr
}
\end{equation}
%%%%%%%%%%
where $\delta v \equiv 2 \delta x \partial_{i} v^{i}$, $\delta x (= \Delta x = 
\Delta y = \Delta z)$ is the local grid spacing and where $C_{\rm vis}$ is the 
dimensionless parameter.  When evolving the matter equations we limit the 
stepsize $\Delta t$ by the following Courant condition ($\Delta t = {\rm min} 
(0.3, C_{\rm dyn} / \sqrt{\rho^{*}_{\rm max}}) \Delta x$), where the second 
term represents dynamical time of a collapsing star.  We choose the dimensionless 
parameter $C_{\rm dyn} \approx 0.01$.

\begin{figure}
\figurenum{1}
\epsscale{1.05}
\plotone{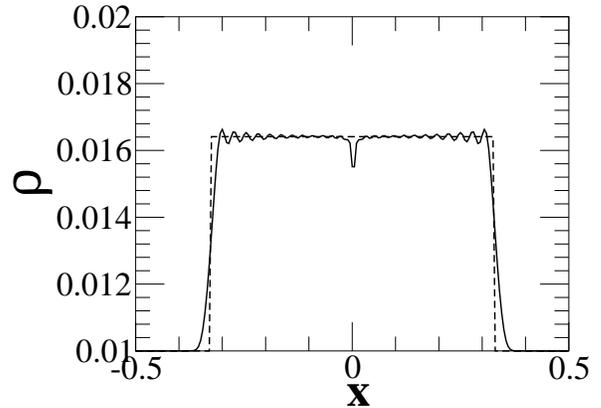}
\caption[f01.eps]{
Comparison between numerical and analytical results in a
one-dimensional relativistic wall shock problem at $t/M = 1.0$
(where the fluid flow is aligned with the $x$-axis).  Solid and dashed
lines represent analytic and numerical results, respectively.  For
this simulation we chose $\Gamma = 1.33$, $\rho^{(0)} = 1.00 \times
10^{-2}$ with
a grid space $\delta x = 5 \times 10^{-3}$ and $v_{0} = 0.200$.
\label{fig:wshock}
}
\end{figure}

We have tested the ability of our code to resolve shocks by performing a 
wall-shock problem, in which two phases of a fluid collide.  In  Fig. 
\ref{fig:wshock} we compare numerical results with the analytic solution for 
initial velocities that are similar to those found in our simulations below.  
With $C_{\rm vis} = 3$ we find good agreement, and set this value to simulate
the catastrophic collapse in \S \ref{sec:jm2}.

%%%%%%%%%%%%%%%%%%%%%%%%%%%%%%%%%%%%%%%%%%%%%%%%%
% Figure 1
%%%%%%%%%%%%%%%%%%%%%%%%%%%%%%%%%%%%%%%%%%%%%%%%%
%%%%%%%%%%
%%%%%%%%%%

\begin{figure}
\figurenum{2}
\epsscale{1.05}
\plotone{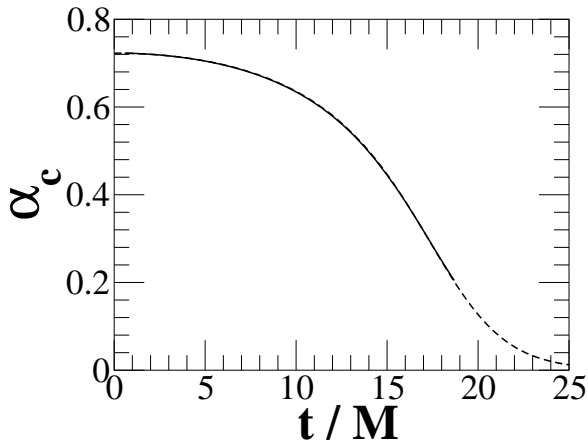}
\caption[f02.eps]{
Comparison of the central lapse between three-dimensional and one-dimensional 
results in Oppenhymer-Snyder collapse.  We start the collapse of a spherical
dust from $R = 4M$.  Solid and dashed lines represent the central lapse of 
three-dimensional and one-dimensional results, respectively.  
\label{fig:oscollapse}
}
\end{figure}

%%%%% Oppenheimer-Snyder Collapse %%%%%
Next, we demonstrate a spherical dust collapse in Fig \ref{fig:oscollapse}.  
Note that our conformally flat approximation retains all the nonlinear terms 
to maintain the exact dynamics for a spherical spacetime.  We compare 1D
and 3D results to check whether our 3D code has an 
ability to reproduce a spherical dust collapse.  Note that we construct our 
1D code following Wilson's approach \citep{Wilson} described in Appendix 
\ref{sec:1Dfield}.  We choose the grid size of 1D computation as 5,000, 
while of 3D one as ($101 \times 101 \times 51$).  We find a very good 
agreement of the central lapse between our 1D result and our 
3D conformally flat one within the error of 1\%. We terminate the integration 
of our 3D code at $t/M = 18.6$, since the convergence of the iteration process 
in our elliptic solver becomes significantly slow.

%%%%%%%%%%%%%%%%%%%%%%%%%%%%%%%%%%%%%%%%%%%%%%%%%
% Figure 2
%%%%%%%%%%%%%%%%%%%%%%%%%%%%%%%%%%%%%%%%%%%%%%%%%
%%%%%%%%%%
%%%%%%%%%%

\begin{figure}
\figurenum{3}
\epsscale{1.05}
\plotone{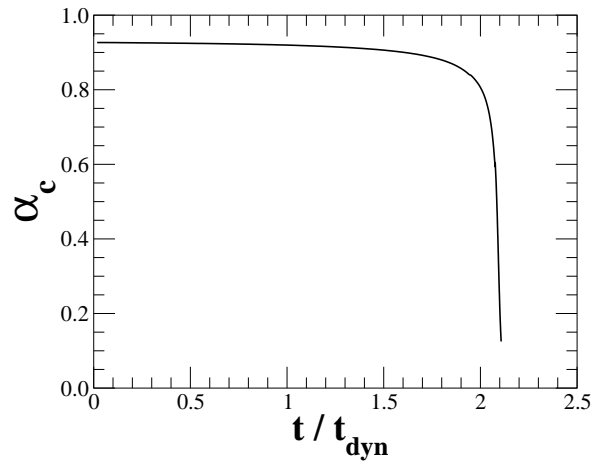}
\caption[f03.eps]{
Central lapse of the spherical collapse (see Table. \ref{tbl:sphinitial}). 
Monotonic decrease of central lapse indicates a prompt collapse
to a spherical BH. 
\label{fig:spalp}
}
\end{figure}

%%%%% Collapse of a Spherical Star %%%%%
Then, we demonstrate the collapse of a spherical star in Fig. \ref{fig:spalp}
(see Table \ref{tbl:sphinitial} for an equilibrium profile).
Since a radially unstable spherical star only collapses to form a spherical 
BH promptly, we should follow the collapse of a spherical star
to check whether our code has an ability to follow the collapse and to find
a signal of BH formation.  We set up the same grid size as we did for a spherical
dust collapse, covering the star with 81 grid points across the diameter.  The 
central lapse of the star decreases monotonically, which means a collapse of a 
spherical star directly forms a spherical BH.  Although we terminate our 
integration around $\alpha_{\rm c} \sim 0.1$, the
fact that 1D computation of the spherical star collapse finds an apparent
horizon at $\alpha_{\rm c} \approx 0.11$ indicates that a BH might form.

%%%%%%%%%%%%%%%%%%%%%%%%%%%%%%%%%%%%%%%%%%%%%%%%%
%   Table 1
%%%%%%%%%%%%%%%%%%%%%%%%%%%%%%%%%%%%%%%%%%%%%%%%%
\begin{deluxetable}{c c c c}
\tablecaption
{Parameters for the initial spherical equilibrium SMS
\label{tbl:sphinitial}}
\tablehead{
\colhead{$\rho_{0}^{\rm max}$\tablenotemark{a}} &
\colhead{$R_{c}$\tablenotemark{b}} &
\colhead{$M$\tablenotemark{c}} &
\colhead{$R_{\rm c} / M$}
}
\startdata
$3.80 \times 10^{-6}$ & $2.49 \times 10^{2}$ & $3.83$ &
$65.0$
\enddata
\tablenotetext{a}{Maximum rest-mass density}
\tablenotetext{b}{Equatorial circumferential radius}
\tablenotetext{c}{Gravitational mass}
\end{deluxetable}
%%%%%%%%%%

%%%%%%%%%%%%%%%%%%%%%%%%%%%%%%%%%%%%%%%%%%%%%%%%%
% Figure 3
%%%%%%%%%%%%%%%%%%%%%%%%%%%%%%%%%%%%%%%%%%%%%%%%%
%%%%%%%%%%
%%%%%%%%%%

%%%%% Stability of a uniformly rotating star %%%%%
Finally, we check the stability of a uniformly rotating star whether our code 
has an ability to determine the radial stability of a star.  Since we 
determine the radial stability of a differentially rotating star in 
\S \ref{sec:jm2}, we should check the sensitivity of our code to 
determine the critical onset of radial instability of uniformly rotating 
stars by comparing the results derived from the turning point method.

To assess the ability of our code to distinguish stable stars from unstable
ones with rotation, we consider an equilibrium sequence of uniformly
rotating stars of the fixed angular momentum $J$ ($J/M^{2} = 0.644$ at
the turning point). While the turning point criterion strictly identifies 
the onset of secular instability, the point of onset of 
dynamical instability nearly coincides with the secular instability point.
We adopt the polytropic index of $n=2.96$,
which is regarded as radiation pressure dominant, SMS sequence,
and grid resolution as in the 
spherical simulations reported above. We decrease the initial pressure
to induce the collapse ($\kappa \rightarrow 0.99 \kappa$).
 
\begin{figure}
\figurenum{4}
\epsscale{1.05}
\plotone{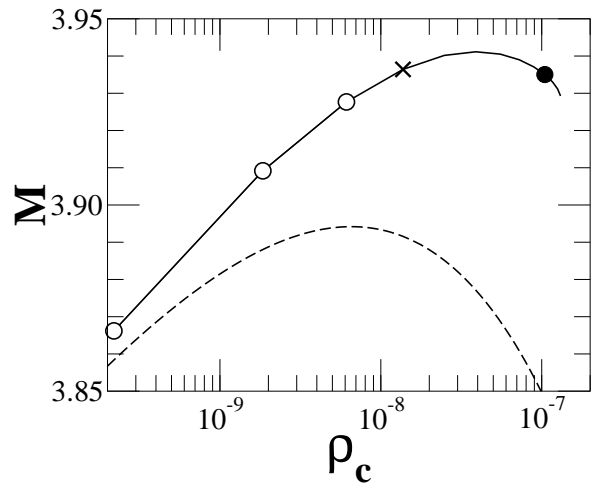}
\caption[f04.eps]{
Probing the dynamical stability of a rotating SMS with $n=2.96$,
$J=10$.
Here, $\rho_{\rm c}$ is the central density of the equilibrium
rotating star.  Filled circles and crosses represent unstable stars,
while open circles represent stable stars according to our dynamical
calculation.  A cross indicates that the star is actually stable
analytically according to the turning point criterion.  
The radii of the 5 marked stars are $R/M = 254$, $421$, $539$, $708$,
and $1579$, where the sequence starts at the right side of the figure
at the  highest central densities.  Note that the solid line shows a
constant $J$ sequence with $J=10$, while the dashed line represents
the spherical equilibrium sequence. With the
adopted grid resolution, our code can distinguish stable stars from 
unstable ones within 0.3\% of the maximum gravitational mass.
\label{fig:roteq}
}
\end{figure}

\begin{figure}
\figurenum{5}
\epsscale{1.05}
\plotone{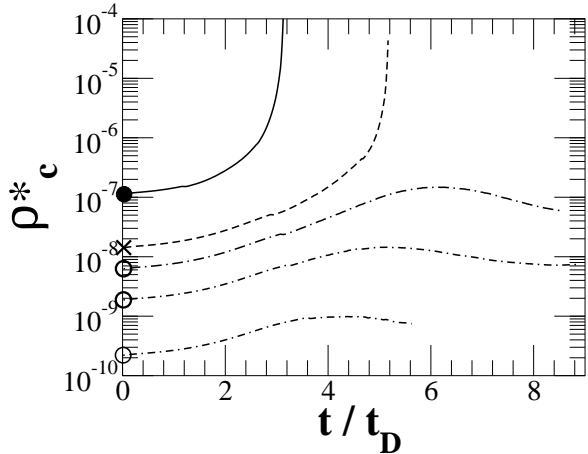}
\caption[f05.eps]{
Evolution of the central densities of the stars plotted in Fig. \ref{fig:roteq}.
Curves are drawn for stars which are unstable both numerically
and according to the turning point criterion (solid),
unstable numerically but stable according to the turning point criterion
(dashed), and stable both numerically and according to the turning point
criterion (dash-dotted).
\label{fig:rotst}
}
\end{figure}

Figure \ref{fig:roteq} summarizes our dynamical stability analysis for
the rotating SMS.  We conclude that with the adopted grid resolution,
our code can distinguish stable rotating stars from unstable ones within
0.3\% of the maximum gravitational mass. Figure \ref{fig:rotst} shows
the evolution of the central density for stable and unstable rotating
stars.

%%%%%%%%%%%%%%%%%%%%%%%%%%%%%%%%%%%%%%%%%%%%%%%%%
% Figure 4
%%%%%%%%%%%%%%%%%%%%%%%%%%%%%%%%%%%%%%%%%%%%%%%%%
%%%%%%%%%%
%%%%%%%%%%

%%%%%%%%%%%%%%%%%%%%%%%%%%%%%%%%%%%%%%%%%%%%%%%%%
% Figure 5
%%%%%%%%%%%%%%%%%%%%%%%%%%%%%%%%%%%%%%%%%%%%%%%%%
%%%%%%%%%%
%%%%%%%%%%

%%%%%%%%%%%%%%%%%%%%%%%%%%%%%%%%%%%%%%%%%%%%%%%%%
%%%%%%%%%%%%%%%%%%%%%%%%%%%%%%%%%%%%%%%%%%%%%%%%%
%%%
%%%   Section IV : Numerical Results
%%%
%%%%%%%%%%%%%%%%%%%%%%%%%%%%%%%%%%%%%%%%%%%%%%%%%
%%%%%%%%%%%%%%%%%%%%%%%%%%%%%%%%%%%%%%%%%%%%%%%%%
%%%%%%%%%%%%%%%%%%%%
\section{Collapse of a differentially rotating supermassive star}
\label{sec:jm2}
%%%%%%%%%%%%%%%%%%%%
Here we explain our initial data sets for collapsing stars.  Since we are 
interested in a collapsing star of $J/M^{2} \sim 1$, we have three 
requirements to construct differentially rotating equilibrium stars.

The first requirement is to construct radially unstable differentially 
rotating stars.  The critical $\Gamma$ to the onset of radial instability 
for slow rotation and weak gravitational field is described analytically as 
\citep{CL,ST83}
%%%%%
\begin{eqnarray}
\Gamma_{\rm crit} = \frac{4}{3} + 2.25 \frac{M}{R} 
- \frac{2}{9} \frac{\Omega^{2} I}{W}
,
\label{eqn:critG}
\end{eqnarray}
%%%%%
where $I$ is the inertia momenta of the star.  Note that relativistic 
gravitation unstabilizes the star, while rotation, which produces centrifugal 
force, stabilizes the star.  Although we take into account the differential 
rotation in the equilibrium star, this criterion (eq. [\ref{eqn:critG}]) may 
indicate the appropriate direction to choose a parameter sets of the 
initial condition of the rotating 
collapsing star.  From eq. (\ref{eqn:critG}), we should at least choose soft 
equation of state to induce the collapse and set an $n=3$ polytropic star,
an SMS sequence of the star.

The second requirement is to construct a star which holds 
$J/M^{2} \approx 1$.  From 
the critical onset of radial instability in the equilibrium star, a uniformly 
rotating SMS takes the maximum of $J/M^{2} \sim 0.9$ when $R/M \sim 600$ 
\citep{BS99,Shibata04}.  To verify the nature of cosmic censor, we should at least go 
beyond $J/M^{2} \gtrsim 1$.  The main restriction to hold large $J/M^2$ comes 
from the mass shedding limit of the star.  Therefore, to construct a star of 
$J/M^{2} > 1$ in radially unstable branch, differential rotation of the star 
is required in SMS sequence.

The last requirement is high degree of differential rotation.
It is indeed one path to construct high degree of differential rotation
in quasi-static evolution, and reach the onset of radial instability.

To summarize, we need soft equation of state and high degree of differential
rotation, and choose an $n=3$ polytropic star (SMS sequences) and
the degree of differential rotation as $\Omega_{\rm c} / \Omega_{\rm eq} 
\approx 10$ where we define the rotation profile as
%%%%%%%%%%
\begin{equation}
u^{t} u_{\varphi} = A^2 (\Omega_{c} - \Omega)
.
\end{equation}
%%%%%%%%%% 
In the Newtonian limit ($u^{t} \rightarrow 1$, $u_{\varphi} \rightarrow 
\varpi^2 \Omega$), this rotation law can be written as 
%%%%%%%%%%
\begin{equation}
\Omega = \frac{A^{2} \Omega_{\rm c}}{\varpi^{2} + A^{2}}
,
\end{equation}
%%%%%%%%%%
where $A$ is the degree of differential rotation, $\varpi$ is the cylindrical
radius of the star.  Since $A$ has a dimension of length, we normalize it with
a proper equatorial radius $\bar{R}_{e}$, ($A = \bar{R}_{e} \hat{A}$). 
Hereafter we choose $\hat{A} = 1/3$ to construct relatively a high degree of
differential rotation.
We briefly summarize our method to construct relativistic
rotating equilibrium stars in Appendix \ref{sec:rst}.

We summarize the parameters of our differentially rotating stars at initial 
in Table \ref{tbl:initial}.  We slightly perturb our initial equilibrium 
state according to
%%%%%%%%%%
\begin{eqnarray}
\rho &=& \rho^{\rm (equilibrium)} 
\left( 1 + 
  \delta^{(1)} \frac{x+y}{R_{\rm e}} +
  \delta^{(2)} \frac{x^{2}-y^{2}}{R_{\rm e}^{2}}
\right),
\end{eqnarray}
%%%%%%%%%%
where $\delta^{(1)} = \delta^{(2)} = 10^{-3}$.  We install $m=1$ and
$m=2$  density
perturbation to provide the seed for one-armed spiral and bar formation, 
if the physical
situation should lead to unstable growth.   We adopt a grid size ($201
\times 201 \times 61$), so that the star is initially covered by 161
points across the equatorial diameter. We evolve the rotating SMS up
to the point at which the conformally flat approximation breaks down.

%%%%%%%%%%%%%%%%%%%%%%%%%%%%%%%%%%%%%%%%%%%%%%%%%
%   Table 2
%%%%%%%%%%%%%%%%%%%%%%%%%%%%%%%%%%%%%%%%%%%%%%%%%
\begin{deluxetable*}{c c c c c}
%\tablewidth{21cm}
\tabletypesize{\scriptsize}
\tablecaption
{Parameters for the initial differentially rotating equilibrium
SMSs
\label{tbl:initial}
}
\tablehead{
\colhead{Parameter} & 
\colhead{Model I} & 
\colhead{Model II} & 
\colhead{Model III} & 
\colhead{Model IV} 
}
\startdata
{$R_{p}/R_{e}$\tablenotemark{a}}  & 
$0.600$ & $0.575$ & $0.550$ & $0.500$ 
\\
{$\rho_{0}^{\rm max}$} & 
$3.38 \times 10^{-6}$ & $3.38 \times 10^{-6}$ & 
$3.38 \times 10^{-6}$ & $3.38 \times 10^{-6}$
\\
{$R_{c}$} & 
$3.06 \times 10^{2}$ & $3.12 \times 10^{2}$ & 
$3.18 \times 10^{2}$ & $3.35 \times 10^{2}$
\\
{$\Omega_{\rm c}$  \tablenotemark{b}} &
$1.45 \times 10^{-3}$ & $1.49 \times 10^{-3}$ & 
$1.53 \times 10^{-3}$ & $1.59 \times 10^{-3}$
\\
{$\Omega_{\rm eq}$ \tablenotemark{c}} &
$1.38 \times 10^{-4}$ & $1.42 \times 10^{-4}$ & 
$1.45 \times 10^{-4}$ & $1.51 \times 10^{-4}$
\\
{$M$} & 
$4.69$ & $4.78$ & $4.88$ & $5.10$
\\
{$J$ \tablenotemark{d}} & 
$2.13 \times 10^{1}$ & $2.31 \times 10^{1}$ &
$2.51 \times 10^{1}$ & $2.96 \times 10^{1}$
\\
{$T/W$\tablenotemark{e}} & 
$6.47 \times 10^{-2}$ & $7.02 \times 10^{-2}$ & 
$7.60 \times 10^{-2}$ & $8.80 \times 10^{-2}$ 
\\
{$J/M^{2}$} & 
$0.97$ & $1.01$ & $1.05$ & $1.14$ 
\\
{$R_{c}/M$} &
$6.53 \times 10^{1}$ & $6.52 \times 10^{1}$ &
$6.52 \times 10^{1}$ & $6.57 \times 10^{1}$ 
\\
{Stability} &
Unstable & Unstable & Stable & Stable
\enddata
\tablenotetext{a}{Ratio of the polar proper radius to the 
equatorial proper radius}
\tablenotetext{b}{Maximum rest-mass density}
\tablenotetext{c}{Equatorial angular velocity at the surface}
\tablenotetext{d}{Angular momentum}
\tablenotetext{e}{Ratio of the rotational kinetic energy to the
gravitational binding energy}
\end{deluxetable*}
%%%%%%%%%%

\begin{figure}
\figurenum{6}
\epsscale{1.05}
\plotone{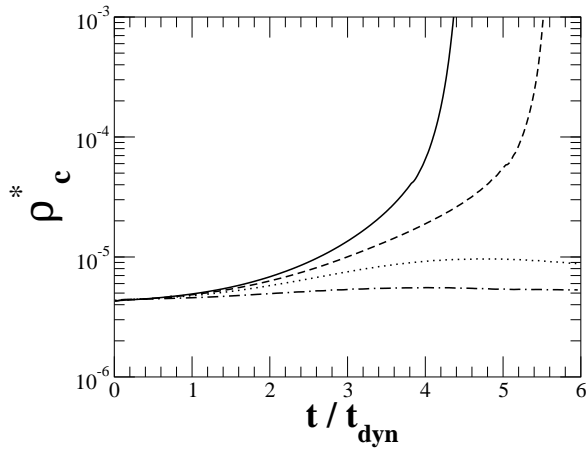}
\caption[f06.eps]{
Evolution of the central density of 4 differentially rotating stars.
Solid, dotted, dashed and dash-dotted line denotes Model I, II, III and
IV (see Table \ref{tbl:initial})
\label{fig:rhomax}
}
\end{figure}

Figure \ref{fig:rhomax} shows the results of radial stability in 4 
differentially rotating stars.  Since we do not have a tool to determine radial 
stability in differentially rotating stars from their equilibrium states (but 
for determining the criterion of secular stability in rotating stars, see \citet{BFG}), the evolution 
is necessary to determine its stability.  The criterion to determine the radial 
stability is as follows.  When the central density of the star grows 
exponentially within a few dynamical time, we determine the star radially 
unstable.  On the other hand, when the central density of the star oscillates 
around its equilibrium state, we determine the star radially stable.  From this 
criterion, Model I and II have an exponential growth of the central density 
that means radially unstable, while Model III and IV have maximum at several 
dynamical times that means radially stable. 

%%%%%%%%%%%%%%%%%%%%%%%%%%%%%%%%%%%%%%%%%%%%%%%%%
% Figure 6
%%%%%%%%%%%%%%%%%%%%%%%%%%%%%%%%%%%%%%%%%%%%%%%%%
%%%%%%%%%%
%%%%%%%%%%

\begin{figure}[t]
\figurenum{7}
\epsscale{1.05}
\plotone{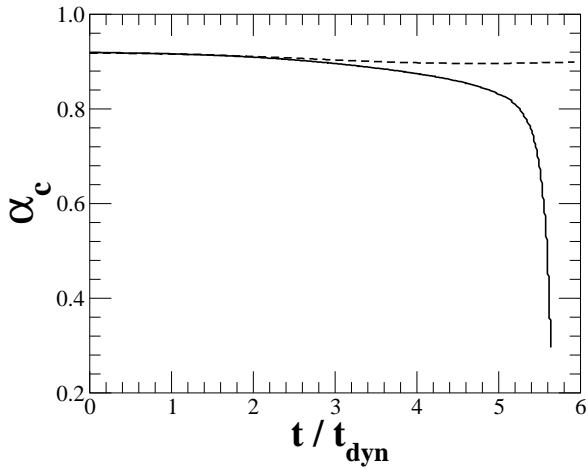}
\caption[f07.eps]{
Evolution of the central lapse of 2 differentially rotating stars.
Solid and dashed line denotes Model II and III, respectively (see Table 
\ref{tbl:initial}).
\label{fig:lapse}
}
\end{figure}

We also show the evolution of central lapse in Fig. \ref{fig:lapse}.  The 
rapid decrease of $\alpha_{\rm c}$ below $0.3$ indicates that a BH is 
likely to 
form.  Model II shows that the central lapse monotonically decreases from 
$\sim 0.9$ to $\sim 0.3$.  This figure shows that we can follow the collapse 
from the regime of Newtonian gravity ($\alpha_{\rm c} \sim 0.9$) to that of 
relativistic gravity ($\alpha_{\rm c} \sim 0.3$).  On the other hand Model III 
shows that the central lapse oscillates around its equilibrium state, and that 
also means the star is radially stable.

%%%%%%%%%%%%%%%%%%%%%%%%%%%%%%%%%%%%%%%%%%%%%%%%%
% Figure 7
%%%%%%%%%%%%%%%%%%%%%%%%%%%%%%%%%%%%%%%%%%%%%%%%%
%%%%%%%%%%
%%%%%%%%%%

\begin{figure*}
\figurenum{8}
\epsscale{1.0}
\plotone{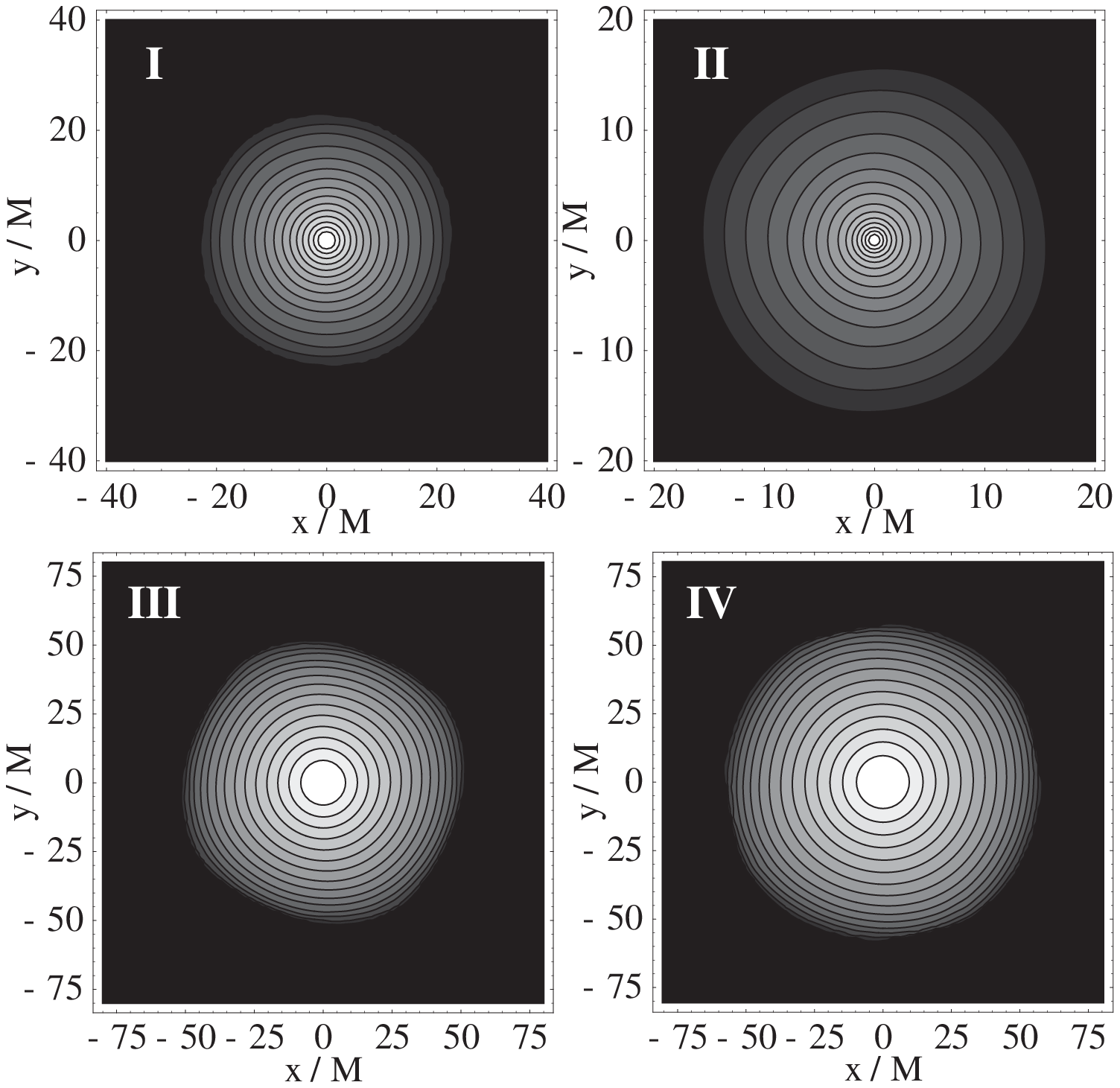}
\caption[f08.eps]{
Final density contour in the equatorial plane of 4 differentially rotating 
stars.  Model I, II, III, 
IV is plotted at the parameter ($t/t_{\rm dyn}$, $\rho^{*}_{\rm max}$)
$=$ ($4.38$, $1.37 \times 10^{-3}$), ($5.62$, $2.16 \times 10^{-2}$), 
($5.61$, $9.13 \times 10^{-6}$), ($5.60$, $5.33 \times 10^{-6}$), respectively.
The contour lines denote densities
$\rho^{*} = \rho^{*}_{\rm max} \times 10^{- 0.267 (16-i)}  (i=1, \cdots, 15)$.
\label{fig:qeq}
}
\end{figure*}

\begin{figure*}
\figurenum{9}
\epsscale{0.70}
\plotone{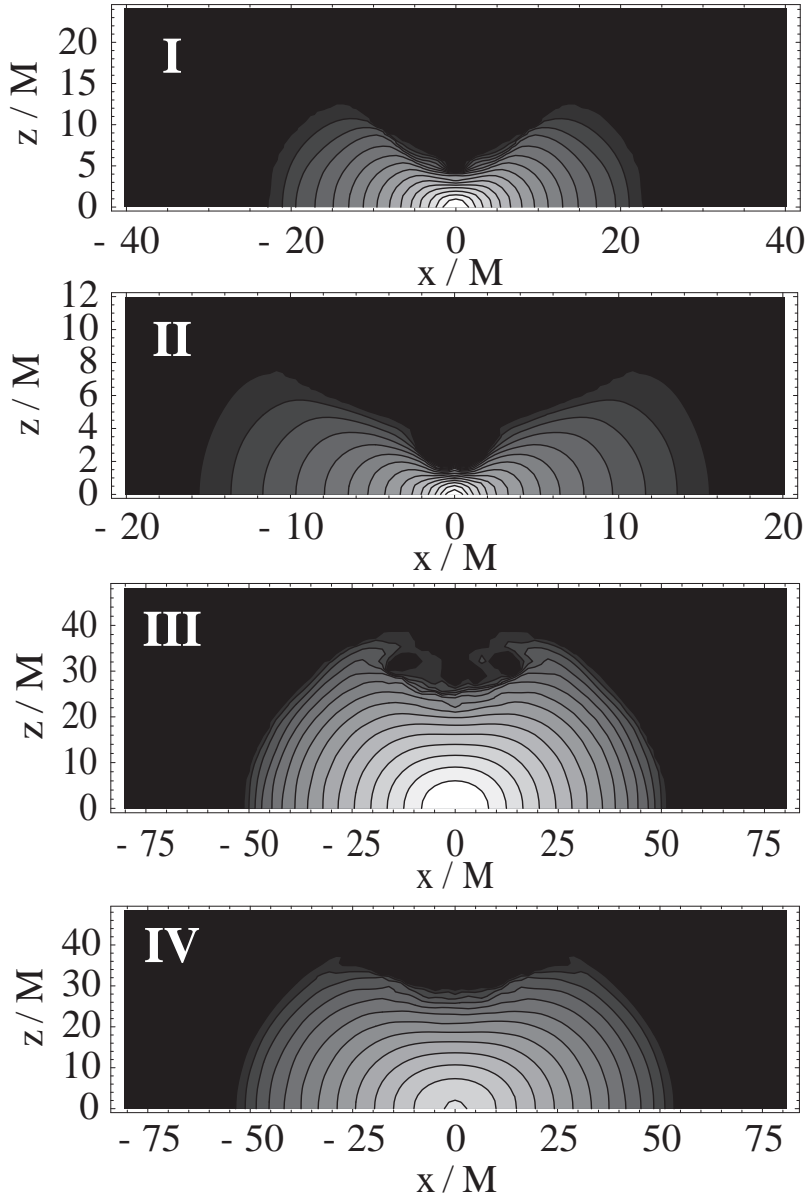}
\caption[f09.eps]{
Final density contour in the meridional plane of 4 differentially rotating 
stars.  The time and contour levels are the same as in Fig. \ref{fig:qeq}.
\label{fig:qmd}
}
\end{figure*}

We show the final density snapshots of the stars in the 
equatorial plane (Fig. \ref{fig:qeq}) and in the meridional plane 
(Fig. \ref{fig:qmd}).  Even in the final snapshots of radially unstable 
stars, the collapse is almost axisymmetric.  Also, from the snapshots 
in the meridional plane, the material around the rotational axes collapses 
faster than the surrounding material due to the strong nonlinear 
gravitational field.

%%%%%%%%%%%%%%%%%%%%%%%%%%%%%%%%%%%%%%%%%%%%%%%%%
% Figure 8
%%%%%%%%%%%%%%%%%%%%%%%%%%%%%%%%%%%%%%%%%%%%%%%%%
%%%%%%%%%%
%%%%%%%%%%

%%%%%%%%%%%%%%%%%%%%%%%%%%%%%%%%%%%%%%%%%%%%%%%%%
% Figure 9
%%%%%%%%%%%%%%%%%%%%%%%%%%%%%%%%%%%%%%%%%%%%%%%%%
%%%%%%%%%%
%%%%%%%%%%

Let us now focus on the collapsing star (Model II) to probe 
the final outcome.
We locally define the cylindrical rest mass $m$, angular momentum $j$, specific
angular momentum $j_{s}$, Keplerian angular velocity $\Omega_{\rm K}$ as
%%%%%%%%%%
\begin{eqnarray}
m &=& 
4 \pi \int_{0}^{\infty} dz \int_{0}^{\varpi} d\varpi 
\varpi \rho^{*}
,\\
j &=& 
4 \pi \int_{0}^{\infty} dz \int_{0}^{\varpi} d\varpi 
\varpi h u_{\varphi}
,\\
j_{s} &=& 
4 \pi \int_{0}^{\infty} dz \int_{0}^{\varpi} d\varpi 
\varpi \rho^{*} h u_{\varphi}
,\\
\Omega_{\rm K} &\equiv& \sqrt{\frac{m}{\varpi^{3}}},
\label{eqn:Keplarian}
\end{eqnarray}
%%%%%%%%%%
to indicate the transport of angular momentum and the distribution 
of $j/m^2$.  We assume the axisymmetric collapse to investigate the final 
outcome, since the meridional density snapshots (Fig. \ref{fig:qmd}) 
behave almost axisymmetry.

\begin{figure}
\figurenum{10}
\epsscale{1.05}
\plotone{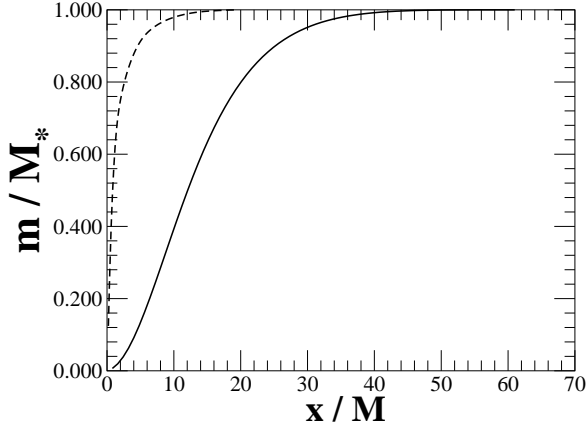}
\caption[f10.eps]{
Mass density profile in the $x$-axes for Model II.
Note that $M_{*}$ is the rest mass.
Solid and dashed line denotes the profile at $t = 0$ and $t = 5.64~t_{\rm
dyn}$, respectively.
\label{fig:mass}
}
\end{figure}

\begin{figure}
\figurenum{11}
\epsscale{1.05}
\plotone{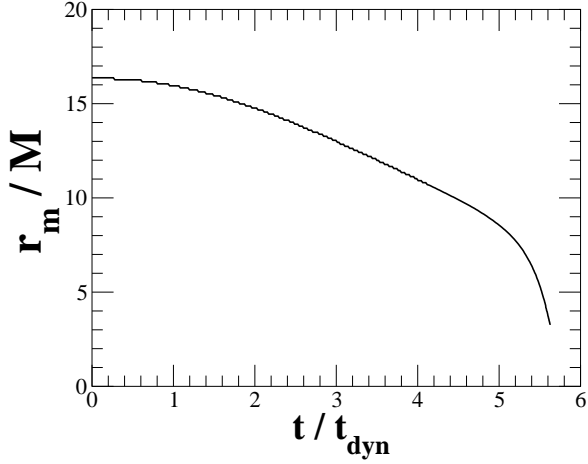}
\caption[f11.eps]{
Mean radius of the star during the collapse for Model II.
\label{fig:rm}
}
\end{figure}

Figure  \ref{fig:mass} shows the concentration of mass density profile 
during the collapse.  As the collapse goes on, the increase ratio of the 
cylindrical mass at the core of the star becomes high.  In fact, $75$\% 
of the cylindrical mass is inside the radius of $r<2M$ at 
$t = 5.64 t_{\rm dyn}$.  We also show the mean radius, defined as
$r_{\rm m} = \sqrt{(\int dv \rho_{*} \varpi^{2}) / M_{*}}$, 
during the collapse in Fig. \ref{fig:rm}.  Note that $M_{*}$ is the rest
mass.  We find that the mean radius monotonically 
decreases, which means that the collapse of the central core is coherent.

%%%%%%%%%%%%%%%%%%%%%%%%%%%%%%%%%%%%%%%%%%%%%%%%%
% Figure 10
%%%%%%%%%%%%%%%%%%%%%%%%%%%%%%%%%%%%%%%%%%%%%%%%%
%%%%%%%%%%
%%%%%%%%%%

%%%%%%%%%%%%%%%%%%%%%%%%%%%%%%%%%%%%%%%%%%%%%%%%%
% Figure 11
%%%%%%%%%%%%%%%%%%%%%%%%%%%%%%%%%%%%%%%%%%%%%%%%%
%%%%%%%%%%
%%%%%%%%%%

\begin{figure}
\figurenum{12}
\epsscale{1.05}
\plotone{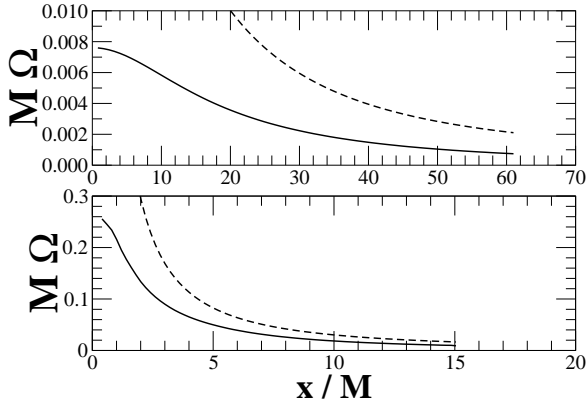}
\caption[f12.eps]{
Angular velocity profile in the $x$-axes for Model II.
Solid and dashed lines denote angular velocity of the star and Keplerian
angular velocity (see eq. [\ref{eqn:Keplarian}] for its definition).  The top
panel shows the snapshots of $t=0$ while the bottom panel of 
$t \sim 5.64~t_{\rm dyn}$.  
\label{fig:omega}
}
\end{figure}

Figure \ref{fig:omega} shows the variation of angular velocity profile 
during the collapse.  Since the cylindrical rest mass density shows a 
coherent collapse, the degree of differential rotation significantly 
increases at the core during the collapse and approaches to the 
``Keplerian'' angular velocity defined in eq. (\ref{eqn:Keplarian}).

%%%%%%%%%%%%%%%%%%%%%%%%%%%%%%%%%%%%%%%%%%%%%%%%%
% Figure 12
%%%%%%%%%%%%%%%%%%%%%%%%%%%%%%%%%%%%%%%%%%%%%%%%%
%%%%%%%%%%
%%%%%%%%%%

\begin{figure}
\figurenum{13}
\epsscale{1.05}
\plotone{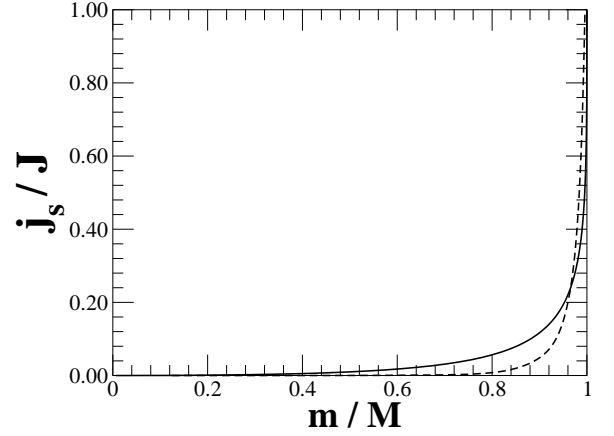}
\caption[f13.eps]{
Specific angular momentum profile as a function of cylindrical mass for Model II.
Solid and dashed line denotes the profile at $t = 0$ and $t = 5.64~t_{\rm
dyn}$, respectively.
\label{fig:jm}
}
\end{figure}

We show the specific angular momentum distribution during the collapse in 
Fig. \ref{fig:jm}.  In a global sense, $J/M^{2}$ is conserved so that 
a final BH should violate cosmic censor, if it forms.  However, we 
find that most of the matter collapses to form a BH, while not a few 
amount of angular momentum stays at the surface area of the star that 
prevents forming a BH of $J/M^{2} > 1$.

%%%%%%%%%%%%%%%%%%%%%%%%%%%%%%%%%%%%%%%%%%%%%%%%%
% Figure 13
%%%%%%%%%%%%%%%%%%%%%%%%%%%%%%%%%%%%%%%%%%%%%%%%%
%%%%%%%%%%
%%%%%%%%%%

\begin{figure}
\figurenum{14}
\epsscale{1.05}
\plotone{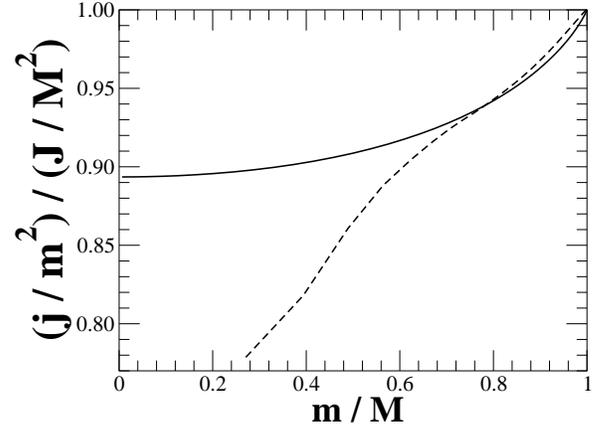}
\caption[f14.eps]{
$j/m^2$ profile as a function of cylindrical mass for Model II.
Solid and dashed line denotes the profile at $t=0$ and $t = 5.64~t_{\rm
dyn}$, respectively.
\label{fig:jm2}
}
\end{figure}

We show the distribution of $j/m^2$ in Fig. \ref{fig:jm2} during the 
collapse to verify cosmic censor.  We find that mass density 
collapses first in the central part and angular momentum remains in 
the surface of the star that prevents forming a BH of $J/M^{2} > 1$.  
Note that the distribution of $j/m^{2}$ is only an indicator to interpret 
the physical cause to prevent BH formation.  Although we should define 
the total gravitational  mass locally to discuss the local distribution 
of $J/M^{2}$, there is no knowing how to define it.  Therefore we use the 
local rest mass instead.

%%%%%%%%%%%%%%%%%%%%%%%%%%%%%%%%%%%%%%%%%%%%%%%%%
% Figure 14
%%%%%%%%%%%%%%%%%%%%%%%%%%%%%%%%%%%%%%%%%%%%%%%%%
%%%%%%%%%%
%%%%%%%%%%

%%%%%%%%%%%%%%%%%%%%%%%%%%%%%%%%%%%%%%%%%%%%%%%%%
%%%%%%%%%%%%%%%%%%%%%%%%%%%%%%%%%%%%%%%%%%%%%%%%%
%%%
%%%   Section V : Discussion
%%%
%%%%%%%%%%%%%%%%%%%%%%%%%%%%%%%%%%%%%%%%%%%%%%%%%
%%%%%%%%%%%%%%%%%%%%%%%%%%%%%%%%%%%%%%%%%%%%%%%%%
%%%%%%%%%%%%%%%%%%%%
\section{Discussion}
\label{sec:Discussion}
%%%%%%%%%%%%%%%%%%%%
We investigate the collapse of differentially rotating SMSs by means of 
hydrodynamic simulations in conformally flat approximation in general 
relativity.  We start our collapse around the onset of radially 
instability at $R/M \sim 65$ to the point where conformally flat 
approximation breaks down.

We find that cosmic censor even holds for a gravitational
collapse of a radially unstable differentially rotating equilibrium SMS of 
$J/M^{2} \gtrsim 1$.  The main reason to prevent formation of a BH of 
$J/M^{2} \gtrsim 1$ is that quite a large amount of angular momentum 
stays at the surface, not core bounce nor shock propagation, bar 
formation in our model.  Note that even a thin disk near the surface of 
the star can hold relatively a large amount of angular momentum if 
the radius is large.  The above conclusion is supported by \citet{SS04} 
in full general relativistic simulation that the criterion of BH 
formation is determined at the central $q_{\rm c}$ 
($\equiv j/m^{2}|_{\varpi \rightarrow 0}$).  In fact, the central 
$q_{\rm c}$ of Model II is $q_{\rm c} \approx 0.89$, which satisfies their 
criterion.
 
The collapse of a differentially rotating, radially unstable SMS is 
coherent and likely leads to formation of an SMBH.  This situation is 
quite similar to the collapse of a uniformly rotating SMS  
\citep[e.g.,][]{SBSS02,SS02}, since $T/W \approx 0.08$ seems still not 
sufficient to 
change the path of the collapse from the one of uniform rotation.  
However, this final outcome may depend on 
the equation of state of the star.  \citet{LR} treated the isothermal 
($\Gamma=1$) collapse of initially homogeneous, uniformly rotating, low 
entropy clouds via smooth particle hydrodynamics (SPH) simulations.  
They found considerable fragmentation into dense clumps, and disk 
formation containing $\sim 5\%$ of the mass. They concluded that a seed 
BH will form at the center and that it likely will grow gradually by accretion
in our model.  Also \citet{Shibata03} treated the collapse of uniformly 
rotating polytropic star from the critical onset in the range of 
$\Gamma \approx 1.5 - 2.5$ and found that the mass of the disk is less 
than $10^{-3}$ of the gravitational mass.

We cannot find any evidence of bar formation nor significant disk formation 
from the rotating collapse prior to BH formation.  The phenomenon of no bar
formation also comes from the fact that mass density collapses first to form a 
BH.  In such case, $T/W$ cannot scale in $R^{-1}$ due to the growth of the 
degree of differential rotation, and as a fact the star of $T/W$ cannot reach 
the dynamical instability point of $\sim 0.27$.  Since the $m=1$ dynamical 
instability takes place when the star has a toroidal structure and soft 
equation of state \citep{CNLB,SBS03}, we cannot find any evidence of toroidal 
structure at the time we stop our integration.  

The rotating SMS collapse is a promising source of burst gravitational waves 
and of quasi-normal mode ringing waves.  The characteristic frequency for 
burst ($f_{\rm burst}$), quasi-normal mode ringing ($f_{\rm QNM}$), 
and the wave amplitude for burst ($h_{\rm burst}$), quasi-normal mode 
ringing ($h_{\rm QNM}$), are \citep{SBSS02}
%%%%%%%%%%
\begin{eqnarray}
f_{\rm burst} &\sim& 
3 \times 10^{-2} 
\left( \frac{10^{6} M_{\odot}}{M} \right) 
\left( \frac{M}{R} \right)^{3/2}
[{\rm Hz}]
,\\
h_{\rm burst} &\sim& 
1 \times 10^{-18} \left( \frac{M}{10^{6} M_{\odot}} \right)
\left( \frac{1 {\rm G pc}}{d} \right)
\left( \frac{M}{R} \right)
,\\
f_{\rm QNM} &\sim&
2 \times 10^{-2} \left( \frac{10^{6} M_{\odot}}{M} \right) {\rm [Hz]}
, \\
h_{\rm QNM} &\sim&
6 \times 10^{-19} \left( \frac{\Delta E_{\rm GW}/M}{10^{-4}} \right)^{1/2}
\left( \frac{2 \times 10^{-2} {\rm [Hz]}}{f_{\rm QNM}} \right)^{1/2}
\nonumber \\
&& \times
\left( \frac{M}{10^{6} M_{\odot}} \right)^{1/2}
\left( \frac{1 {\rm Gpc}}{d} \right)
,
\end{eqnarray}
%%%%%%%%%%
where $d$ is the distance from the observer and $\Delta E_{\rm GW}$ is the 
total radiated energy. We set $R/M = 1$, a characteristic mean radius during
BH formation.  Since the main targets of LISA are gravitational
radiation sources between $10^{-4}$ and $10^{-1}$ Hz, it is possible
that LISA can search for the burst and quasi-normal ringing waves 
accompanying rotating SMS collapse and formation of an SMBH.

%%%%%%%%%%%%%%%%%%%%%%%%%%%%%%%%%%%%%%%%%%%%%%%%%
\acknowledgements
%%%%%%%%%%%%%%%%%%%%%%%%%%%%%%%%%%%%%%%%%%%%%%%%%
We would like to thank an anonymous referee for his/her critical reading 
of our manuscript and constructive suggestions.  
He also thanks Hideki Asada, Silvano Bonazzola, Eric 
Gourgoulhon, Takashi Nakamura, Ken-ichi Nakao, Luciano Rezzolla, Stu 
Shapiro, Nick Streigeolous, Takahiro Tanaka, Koji Ury\=u, Shin Yoshida for 
discussion.  
This work has been supported in part by MEXT Grant-in-Aid for young 
scientists (No. 200200927), Astronomical Data Analysis Center, National 
Astronomical Observatory of Japan, and by MEXT Grant-in-Aid for 21COE 
program at the Department of Physics, Kyoto University.  Numerical 
computations were performed on the NEC SX-5 machine in the Yukawa Institute 
for 
Theoretical Physics, Kyoto University, on the VPP-800 machine in the 
Academic Center for Computing and Media Studies, Kyoto University, and 
on the VPP-5000 machine in the 
Astronomical Data Analysis Center, National Astronomical Observatory of 
Japan.

%%%%%%%%%%%%%%%%%%%%%%%%%%%%%%%%%%%%%%%%%%%%%%%%%
%%%%%%%%%%%%%%%%%%%%%%%%%%%%%%%%%%%%%%%%%%%%%%%%%
%%%
%%%   Appendix
%%%
%%%%%%%%%%%%%%%%%%%%%%%%%%%%%%%%%%%%%%%%%%%%%%%%%
%%%%%%%%%%%%%%%%%%%%%%%%%%%%%%%%%%%%%%%%%%%%%%%%%
\appendix

%%%%%%%%%%%%%%%%%%%%
\section{Relativistic hydrodynamics in Spherically Symmetric Spacetime}
\label{sec:1Dfield}
%%%%%%%%%%%%%%%%%%%%

Here we briefly explain relativistic hydrodynamics in spherically symmetric 
spacetime \citep{Wilson,ST80,BW} that we use in \S~ \ref{sec:CF}.
We can describe the line element of spherically symmetric spacetime 
in the isotropic coordinate as
%%%%%%%%%%
\begin{equation}
ds ^{2} = (- \alpha^{2} + \psi^{4} \beta^{2}) dt^{2} 
+ 2 \psi^{4} \beta dt dr + \psi^{4} (dr^{2} + r^{2} d\Omega),
\end{equation}
%%%%%%%%%%
where $\alpha$ is the lapse, $\beta$ is the shift, $\psi$ is 
the conformal factor.

We use the maximal slicing condition $K=0$, which derives the following
relation to the components of an extrinsic curvature as
%%%%%%%%%%
\begin{equation}
K^{r}_{r} = - 2 K^{\theta}_{\theta} ( = - 2 K^{\phi}_{\phi}).
\end{equation}
%%%%%%%%%%
As a consequence, we only need to consider $K^{r}_{r}$ to construct an 
extrinsic curvature.

The momentum constraint derives the equation of the extrinsic curvature as
%%%%%%%%%%
\begin{equation}
\frac{\partial}{\partial r} ( r \psi^{2} K^{r}_{r} )
= (r \psi^{2})^{3} 8 \pi J_{r}
.
\label{eqn:1DMom}
\end{equation}
%%%%%%%%%%
The integration form of eq. (\ref{eqn:1DMom}) is
%%%%%%%%%%
\begin{equation}
K^{r}_{r} = 
\frac{8 \pi}{(r \psi^{2})^{3}} \int^{r} (r \psi^{2})^{3} J_{r} dr
.
\label{eqn:1DintK}
\end{equation}
%%%%%%%%%%
The restriction of the spatial metric to be conformally flat requires the 
following equation
%%%%%%%%%%
\begin{equation}
2 \alpha K^{r}_{r} = \frac{4}{3} r \partial_{r} \left( \frac{\beta}{r} \right),
\label{eqn:1Dbet}
\end{equation}
%%%%%%%%%%
which gives us an appropriate boundary condition for the extrinsic curvature as 
$O(r^{-3})$.  Also the integration form of eq. (\ref{eqn:1Dbet}) is
%%%%%%%%%%
\begin{equation}
\beta = 
\frac{3}{2} r \int_{r} \frac{\alpha K^{r}_{r}}{r} dr
.
\label{eqn:1Dintb}
\end{equation}
%%%%%%%%%%

The Hamiltonian constraint and the trace of the evolution equation guide
%%%%%%%%%%
\begin{eqnarray}
\frac{1}{r^{2}} \partial_{r} ( r^{2} \partial_{r} \psi )
=  - 2 \pi \psi^{5} \rho_{\rm H} - \frac{3}{16} \psi^{5} (K^{r}_{r})^{2}
\label{eqn:1DHam}
,\\
\frac{1}{r^{2}} \partial_{r} [ r^{2} \partial_{r} ( \alpha \psi ) ]
=  2 \pi \alpha \psi^{5} (\rho_{\rm H} + 2 S) 
+ \frac{21}{16} \alpha \psi^{5} (K^{r}_{r})^{2}
\label{eqn:1Dev}
,
\end{eqnarray}
%%%%%%%%%%
with the boundary conditions
%%%%%%%%%%
\begin{eqnarray}
\psi &=& 1 + \frac{M}{2 r} + o(r^{-3})
,\\
\alpha \psi &=& 1 + \frac{M}{r} + o(r^{-3})
,
\end{eqnarray}
%%%%%%%%%%
at the grid edge and
%%%%%%%%%%
\begin{equation}
\partial_{r} \alpha = 
\partial_{r} \psi = 0
,
\end{equation}
%%%%%%%%%%
at the center ($r=0$).

Therefore, we can determine 4 unknown variables ($K^{r}_{r}$, $\psi$, $\alpha
\psi$, $\beta$) with 4 equations (eqs. [\ref{eqn:1DintK}], [\ref{eqn:1Dintb}], 
[\ref{eqn:1DHam}], [\ref{eqn:1Dev}]). \footnote{We choose 
$\hat{A}_{r}^{r} = \psi^{6} K_{r}^{r}$ instead of $K^{r}_{r}$ from a 
computational point of view (See \S \ref{subsec:NT}.).}

Matter equations can be written as 
%%%%%%%%%%
\begin{eqnarray}
&&
\frac{\partial (r^2 \rho_{*})}{\partial t}+
\frac{\partial (r^2 \rho_{*} v^{r})}{\partial r} = 
0
,\\
&&
\frac{\partial (r^2 e_{*})}{\partial t}+
\frac{\partial (r^2 e_{*} v^{r})}{\partial r} = 
0
,\\
&&
\frac{\partial(r^{2} \rho_{*} \tilde u_{r})}{\partial t}
+ \frac{\partial (r^{2} \rho_* \tilde u_{r} v^{r})}{\partial r} 
=
- r^{2} \alpha \psi^{6} P_{,r} 
- \rho_{*} \alpha \tilde u^{t} \alpha_{,r} 
\nonumber \\
&& \hspace{2cm}
+ r^{2} \rho_{*} \tilde u_{r} \beta^{r}_{~,r}
+ r^{2} \frac{2 \rho_{*} \tilde u_{r} \tilde u_{r}}{\psi^{5} \tilde u^{t}} 
\psi_{,r}
.
\end{eqnarray}
%%%%%%%%%%
We have not taken into account artificial viscosity, since shock does not seem
to play an important role in a spherical dust collapse and the collapse of 
a spherical star.  In \S~\ref{sec:ctest}, we 
adopt our 1D relativistic hydrodynamic code to a spherical dust collapse by 
setting $e_{*}$ and $P_{*}$ $\ll$ equilibrium variables of them in typical
polytropic index.

We briefly mention a method to compute an apparent horizon in 1D 
\citep{ST80,NOK87}.  An apparent horizon is defined as an outermost 
marginally trapped surface whose future-directed outgoing null geodesics
have zero expansion $\Theta$ \citep{HE}.  The outgoing null vector $k^{\mu}$ 
can be expressed as $k^{\mu} = (n^{\mu} + s^{\mu}) / \sqrt{2}$, where
$n^{\mu}$ is unit normal of the spacelike hypersurface defined as
$n^{\mu} =$($-\alpha$, 0, 0, 0), $s^{\mu}$ is the out-directed spacelike vector
orthogonal to the surface.  The projection tensor $h_{\mu \nu}$ and 
the extrinsic curvature $K_{ij}$ are described as 
%%%%%%%%%%
\begin{eqnarray}
h_{\mu \nu} &=& g_{\mu \nu} + n_{\mu} n_{\nu},\\
K_{ij} &=& h_{i}^{\mu} h_{j}^{\nu} \nabla_{\mu} n_{\nu}.
\end{eqnarray}
%%%%%%%%%%
The expansion $\Theta$ is then written as
%%%%%%%%%%
\begin{equation}
\Theta \equiv \nabla_{\mu} k^{\mu} = 
\nabla_{i} s^{i} + K_{ij} s^{i} s^{j} - K.
\label{eqn:MTS}
\end{equation}
%%%%%%%%%%
For spherically symmetric spacetime, we can choose the spacelike vector 
$s^{i}$ as $s^{i} = (\psi^{-2}, 0, 0)$, and then the radius $r_{\rm AH}$
of an apparent horizon satisfies the following equation,
%%%%%%%%%%
\begin{equation}
\left.
1 + 2 r \frac{\psi_{,r}}{\psi} + \frac{1}{2} r \psi^{2} K_{r}^{r} 
\right|_{r = r_{\rm AH}}
= 0.
\end{equation}
%%%%%%%%%%
Note that an event horizon always lies outside an apparent horizon, and both 
horizons coincide with each other when the spacetime is stationary.
Also an event horizon always forms before the time that an apparent 
horizon does \citep{HE}.

%%%%%%%%%%%%%%%%%%%%
\section{Construction of relativistic rotating equilibrium stars}
\label{sec:rst}
%%%%%%%%%%%%%%%%%%%%

Here we summarize our method to construct rotating relativistic 
equilibrium stars, which is based on \citet{KEH,CST92,CST94}
\citep[see also][for a historical review]{Stergioulas03}.

First we focus on how to solve the relativistic Euler equation.
The equation can be described in axisymmetric spacetime as
%%%%%%%%%%
\begin{equation}
\frac{h_{,j}}{h} - \frac{u^{t}_{,j}}{u^{t}} + u^{t} u_{\varphi} \Omega_{,j} 
= 0
,
\label{eqn:REuler}
\end{equation}
%%%%%%%%%%
where $h \equiv ( 1 + \varepsilon + P/\rho_{0} )$ is a specific enthalpy.
We also assume specific type of rotation law as
%%%%%%%%%%
\begin{equation}
u^{t} u_{\varphi} = A^{2} (\Omega_{\rm c} - \Omega)
,
\label{eqn:RLaw}
\end{equation}
%%%%%%%%%%
where $\Omega_{c}$ is the central angular velocity, in order to integrate 
eq. (\ref{eqn:REuler}).  The Bernoulli's equation is driven by integrating 
the relativistic Euler equation (\ref{eqn:REuler}) as
%%%%%%%%%%
\begin{equation}
H - K + R = C
,
\label{eqn:Bernoulli}
\end{equation}
%%%%%%%%%%
where
%%%%%%%%%%
\begin{eqnarray}
H &\equiv& \int \frac{dh}{h} = \ln h = \ln [1+(n+1)q]
,\\
K &\equiv& \int \frac{du^{t}}{u^{t}} = \ln u^{t} 
= - \frac{1}{2} \ln 
[\alpha^{2} - \psi^{4} [(\beta^{x} - \Omega y)^{2} + 
(\beta^{y} + \Omega x)^{2}]]
,\\
R &\equiv& \int u^{t} u_{\varphi} d\Omega 
= -\frac{1}{2} A^{2} (\Omega_{c} - \Omega)^{2}
,
\end{eqnarray}
%%%%%%%%%%
where $q \equiv P/\rho_{0}$\footnote{We only use $q \equiv P/\rho_{0}$ in 
this section.}.  Note that we have already assumed polytropic 
equation of state $P = \rho_{0}^{\Gamma}$ where $\Gamma = 1 + 1/n$, $n$ is 
a polytropic index.  Therefore, we can describe the matter distribution 
equation by using eq. (\ref{eqn:Bernoulli}) and the rotation law 
(eq. [\ref{eqn:RLaw}]) as
%%%%%%%%%%
\begin{eqnarray}
q 
&=&  \frac{1}{n+1} 
\left[ 
  \frac{C \exp[A^{2}(\Omega_{c} - \Omega)^{2} / 2]}{\sqrt{u^{t}}} - 1
\right]
\nonumber \\
&=& \frac{1}{n+1} 
\left[ 
  \frac{C \exp[A^{2}(\Omega_{c} - \Omega)^{2} / 2]}
       {\sqrt{\alpha^{2} - \psi^{4} [(\beta^{x} - \Omega y)^{2} + 
         (\beta^{y} + \Omega x)^{2}]}} - 1
\right]
\label{eqn:qdis}
,\\
A^{2} (\Omega_{c} - \Omega) &=& 
u^{t} u_{\varphi} =
\frac{\psi^{4} [x (\beta^{y} + \Omega x) - y (\beta^{x} - \Omega y)]}
{\alpha^{2} - 
  \psi^{4} [(\beta^{x} - \Omega y)^{2} + (\beta^{y} + \Omega x)^{2}]}
.
\label{eqn:rotlaw}
\end{eqnarray}
%%%%%%%%%%
Note that we assume conformally flat spacetime to derive the final
part of the equality in eqs. (\ref{eqn:qdis}) and (\ref{eqn:rotlaw}).

From the computational point of view, we introduce nondimensional quantities
rescaled in the equatorial proper radius of the star ($R_{e}$) as
%%%%%%%%%%
\begin{equation}
\begin{array}{c c c}
\hat{x} = x / R_{e}
, &
\hat{y} = y / R_{e}
, &
\hat{z} = z / R_{e}
, \\
\hat{\Omega} = R_{e} \Omega
, &
\hat{A} = A / R_{e}
. & 
\hat{\triangle} = R_{e}^{2} \triangle
.
\end{array}
\end{equation}
%%%%%%%%%%
We also rescale the lapse and conformal factor using the nature of 
scale free in Newtonian gravity as
%%%%%%%%%%
\begin{equation}
\hat{\alpha} = \alpha^{1 / R_{e}^{2}},
\hat{\psi} = \psi^{-1/(2 / R_{e}^{2})}.
\end{equation}
%%%%%%%%%%

To determine the matter distribution and rotation profile of the next 
iteration step, 
we have to determine five unknown variables ($R_{e}$, $C$, $\Omega_{c}$, 
$\Omega_{e}$, $\Omega_{\rm max}$) from five equations.  Therefore, we evaluate
eqs. (\ref{eqn:qdis}) and (\ref{eqn:rotlaw}) at three locations, the point of 
the maximum rest mass density, that of polar surface, and that of the 
equatorial surface.   Note that eq. (\ref{eqn:rotlaw}) becomes an identity 
at polar surface of the star, and thus we have five equations to be solved.  
We use Newton-Rapson method \citep{NR} to solve these equations.  Once we 
determine 5 unknown variables, we solve eq. (\ref{eqn:rotlaw}) using 
Newton-Rapson method to determine  the rotation profile.  After that we can 
determine the matter distribution (eq. [\ref{eqn:qdis}]).

Next we briefly summarize the gravitational field equations normalized by 
proper radius as follows.
%%%%%%%%%%
\begin{eqnarray}
\hat{\Delta} B_{i} &=& 8 \pi R_{e}^{2} \psi^{6} J_{i} \equiv 4 \pi S_{B_{i}},
\label{eqn:ApCFBx}
\\
\hat{\Delta} \chi  &=& - 8 \pi R_{e}^{2} \psi^{6} J_{i} x^{i} \equiv 4 \pi S_{\chi},
\\
\hat{\Delta} \psi  &=& 
- 2 \pi R_{e}^{2} \psi^{5} \rho_{\rm H} - 
\frac{1}{8} \psi^{-7} \hat{A}_{ij} \hat{A}^{ij}
\equiv 4 \pi S_{\psi},
\\
\hat{\Delta} (\alpha \psi) &=& 
2 \pi R_{e}^{2} \alpha \psi (\rho_{\rm H} + 2 S) +
\frac{7}{8} \alpha \psi^{-7} \hat{A}_{ij} \hat{A}^{ij}
\equiv 4\pi S_{\alpha\psi}, 
\\
\hat{\Delta} P_{i} &=& 4 \pi R_{e}^{2}\alpha \hat{J}_{i} \equiv 4 \pi S_{P_{i}},
\\
\hat{\Delta} \eta  &=& -4 \pi R_{e}^{2} \alpha \hat{J}_{i} x^{i} \equiv 4 \pi S_{\eta}
\label{eqn:ApCFeta}
.
\end{eqnarray}
%%%%%%%%%%
We choose the same boundary conditions as we choose in our evolution code
(see \S~\ref{sec:CF}).  Note that we construct the star in 3D using octant 
symmetry to adopt our initial data to our 3D evolution code smoothly.

To summarize, we first solve the gravitational field equations (eqs.
[\ref{eqn:ApCFBx}] -- [\ref{eqn:ApCFeta}]).  Next, we determine 5
unknown quantities ($R_{e}$, $C$, $\Omega_{c}$, $\Omega_{e}$, 
$\Omega_{\rm max}$) from 5 equations (eqs. [\ref{eqn:qdis}] and 
[\ref{eqn:rotlaw}]) using Newton-Rapson method.  
Finally we determine the matter distribution and rotation profile for the 
next iteration step.  We continue this iteration cycle till $R_{e}$ 
converges; i.e. $R_{e}$ goes bellow the error of $10^{-5}$.  We also 
check the error rate of physical quantities such as $M$ and $J$ and find 
that they are in fact below the error rate of $10^{-4}$.

%%%%%%%%%%%%%%%%%%%%%%%%%%%%%%%%%%%%%%%%%%%%%%%%%
%%%%%%%%%%%%%%%%%%%%%%%%%%%%%%%%%%%%%%%%%%%%%%%%%
%%%
%%%   References
%%%
%%%%%%%%%%%%%%%%%%%%%%%%%%%%%%%%%%%%%%%%%%%%%%%%%
%%%%%%%%%%%%%%%%%%%%%%%%%%%%%%%%%%%%%%%%%%%%%%%%%

\end{document}